\documentclass[12pt,preprint]{aastex}

\usepackage{graphicx}
\usepackage{amsfonts}
\usepackage{amsmath}
\usepackage{amssymb}
\usepackage{amsthm}
\usepackage{pdflscape}
\usepackage{wasysym}

\newcommand{\etal}{et~al.\ }

\newcommand{\flux}{\hbox{erg~cm$^{-2}$~s$^{-1}$}}

\newcommand{\be}{\begin{equation}}
\newcommand{\ee}{\end{equation}}
\newcommand{\ba}{\begin{eqnarray}}
\newcommand{\ea}{\end{eqnarray}}

\newcommand{\chandra}{{\emph{Chandra}}}

\newcommand{\xmm}{\emph{XMM-Newton}}

\newcommand{\swift}{\emph{Swift}}

\newcommand{\simgt}{\lower 2pt \hbox{$\, \buildrel {\scriptstyle >}\over {\scriptstyle\sim}\,$}}
\newcommand{\simlt}{\lower 2pt \hbox{$\, \buildrel {\scriptstyle <}\over {\scriptstyle\sim}\,$}}
\newcommand{\ls}{\lower 2pt \hbox{$\;\scriptscriptstyle \buildrel<\over\sim\;$}}
\newcommand{\gs}{\lower 2pt \hbox{$\;\scriptscriptstyle \buildrel>\over\sim\;$}}

\newcommand{\marc}{$^{\prime}\!\!.$}

\usepackage{color}

\begin{document}

\def\arcsec{$^{\prime\prime}$}
\def\arcmin{$^{\prime}$}
\def\degr{$^{\circ}$}

\title{The \swift\ AGN and Cluster Survey. II. Cluster Confirmation with SDSS Data}

\author{Rhiannon D. Griffin\altaffilmark{1}, Xinyu Dai\altaffilmark{1}, Christopher S. Kochanek\altaffilmark{2}, Joel N. Bregman\altaffilmark{3}} 

\altaffiltext{1}{Homer L. Dodge Department of Physics and Astronomy,
University of Oklahoma, Norman, OK, 73019;
Rhiannon.D.Griffin-1@ou.edu, xdai@ou.edu}
\altaffiltext{2}{Department of Astronomy, Ohio State University, Columbus, OH 43210; ckochanek@astronomy.ohio-state.edu}
\altaffiltext{3}{Department of Astronomy, University of Michigan, Ann Arbor, MI 48109; jbregman@umich.edu}

\begin{abstract}
We study 203 (of 442) \swift\ AGN and Cluster Survey extended X-ray sources located in the SDSS DR8 footprint to search for galaxy over-densities in three dimensional space using SDSS galaxy photometric redshifts and positions near the \swift\ cluster candidates.
We find 104 \swift\ clusters with a $>3\sigma$ galaxy over-density. 
The remaining targets are potentially located at higher redshifts and require deeper optical follow-up observations for confirmation as galaxy clusters.  
We present a series of cluster properties including the redshift, BCG magnitude, BCG-to-X-ray center offset, optical richness, and X-ray luminosity.
We also detect red sequences in $\sim85\%$ of the 104 confirmed clusters.
The X-ray luminosity and optical richness for the SDSS confirmed \swift\ clusters are correlated and follow previously established relations.  
The distribution of the separations between the X-ray centroids and the most likely BCG is also consistent with expectation.
We compare the observed redshift distribution of the sample with a theoretical model, and  
find that our sample is complete for $z \lesssim 0.3$ and is still 80\% complete up to $z \simeq 0.4$, consistent with the SDSS survey depth.
These analysis results suggest that our \swift\ cluster selection algorithm has yielded a statistically well-defined cluster sample for further studying cluster evolution and cosmology.
We also match our SDSS confirmed \swift\ clusters to existing cluster catalogs, and find 42, 23 and 1  matches in optical, X-ray and SZ catalogs, respectively, so the majority of these clusters are new detections.  
\end{abstract}


\section{Introduction}
Our universe is organized in a cosmic web on megaparsec (Mpc) scales, with filaments, voids and massive over-densities of matter.  These most massive peaks in the large scale matter density are traced by galaxy clusters, the largest gravitationally bound structures in the universe (e.g., \citealt{bahcall,formofgal}). 
Large samples of clusters together with subsequent mass and redshift estimates, allow us to constrain the cluster mass function and thus place improved constraints on important cosmological parameters such as $\sigma_8, \Omega_m$, and $w$ (see Allen~\etal~2011 for a recent review) as well as studying cluster evolution across cosmic time.  

Several methods are employed to discover galaxy clusters.  Optical identification produces the largest cluster catalogs by far, and different algorithms focus on different aspects of optical properties, such as spatial galaxy over-densities (e.g., \citealt{ opt1, opt2, kochanek}), the red sequence (e.g., \citealt{nilo, red1, red2}), or the brightest cluster galaxy (BCG) (e.g., \citealt{maxbcg}).  
Optical identification schemes usually suffer from projection effects created by observing a three-dimensional object in a two-dimensional plane.  
However, optical surveys play a crucial role in measuring cluster redshifts \citep[e.g.][]{xmmlss}.  More recent methods include the Sunyaev-Zel'dovich (SZ) effect and gravitational lensing.  The SZ effect is caused by cosmic microwave background (CMB) photons inverse Compton scattering off the high energy electrons of the intracluster medium (ICM) (e.g., \citealt{sz2, sz1}). This can be seen as a distortion in the shape of the CMB spectrum and is used to follow-up known clusters as well as for new cluster surveys (e.g., \citealt{ sz1, sz3, sz4, sz5, sz6, sz7}).  
Gravitational lensing provides a direct measure of the mass of a cluster and is observable through the deflection, shearing, and magnification of background sources (e.g., \citealt{gravlens1, gravlens7}).
Lensing and cosmic shear surveys (e.g., \citealt{gravlens2, gravlens3, gravlens5, gravlens6, gravlens7}) trace the large scale structure (LSS), provide independent cluster mass estimates, map the dark matter within, and place independent constraints with improved calibration on cosmological parameters (e.g. \citealt{gravlens1, gravlens8}) and potentially could be a useful detection method in future surveys (e.g., \citealt{gravlens4}), as demonstrated by the Deep Lens Survey \citep{gravlens9}. 

Galaxy clusters also appear as extended sources in the X-ray sky.  
Hot electrons in the ICM interact with protons and atomic nuclei to cause plasma emission in the X-ray regime (e.g., \citealt{formofgal}).
X-ray selected cluster surveys have been performed with many different combinations of survey depth and area, such as NORAS (Northern ROSAT All-Sky Survey, \citealt{noras}), 400SD (400 Square Degree Survey, \citealt{400sd}), and REFLEX (ROSAT-ESO Flux-Limited X-Ray Survey, \citealt{reflex}).  These studies and more are included in the  MCXC catalog\footnote{Meta-Catalog of X-ray detected Clusters of galaxies: \texttt{http://heasarc.gsfc.nasa.gov/W3Browse/all/mcxc.html}} \citep{mcxc}, a compilation of X-ray selected galaxy clusters and their characteristics.  The MCXC catalog includes catalogs based on publicly available ROSAT All Sky Survey and serendipitous observations. 
Recent examples of \xmm\ and/or \chandra\ selected cluster and group surveys include the ChaMP (\chandra\ Multiwavelength Project) Serendipitous Galaxy Cluster Survey \citep{barkhouse}, galaxy groups in the Extended \chandra\ Deep Field South \citep{finoguenov15}, the \textit{XMM} Cluster Survey \citep{mehrtens}, and the \xmm\ Wide-Field Survey in the COSMOS Field \citep{finoguenov}.

X-ray identification has several advantages over optical.  First, X-ray observations of clusters suffer less from the projection effects that limit most optical surveys. Because the X-ray emissivity is proportional to the square of the electron density, it provides good contrast over any background (e.g., \citealt{voit}).  
Furthermore, there are several scaling relations based on X-ray data that galaxy clusters are known to follow.
For example, the X-ray luminosity versus mass ($L_X$$-M$) relation is much tighter than the optical richness to mass relation and is a more accurate mass determination method (e.g., \citealt{noras,voit}). 
One disadvantage to the X-ray identification of galaxy clusters is that it is difficult to detect high redshift clusters because of the $(1+z)^{-4}$ dependence of surface brightness on redshift.  Thus, shallow X-ray surveys only detect the core region of high redshift clusters, which increasingly resemble point-like sources.  Also, low redshift galaxies can be extended X-ray sources and may appear as false positives in X-ray cluster surveys (e.g., \citealt{xmmlss}).  Unfortunately, it is also difficult to get accurate measurements of redshifts from X-ray data alone.  

Therefore using X-ray and optical data in a combined program, like the one presented in this paper, is ideal for determining cluster characteristics. 
Since the mass of a galaxy cluster is dominated by dark matter, their total masses are difficult to measure directly.  
Thus, mass estimates are obtained by correlating cluster mass with easily observable quantities that include X-ray luminosity, richness (number of member galaxies), temperature and velocity dispersions (e.g., \citealt{voit,baxnosocs}). 
Proving good correlations between independent mass estimates and further constraining them is imperative so that reliable measurements of the cluster mass function can be obtained. This allows us to investigate its evolution in time and thus better study the formation and evolution of structures and improve constraints on cosmological parameters (e.g., \citealt{formofgal,baxnosocs}).
For example, in this combined program we correlate the X-ray bolometric luminosity and the optical richness ($N_{opt}$), comparing our results with other studies (see Section \ref{subsec:xlumvsopt}).

The \swift\ gamma-ray burst (GRB) fields provide 125 deg$^2$ of serendipitous soft X-ray observations.
Several other groups are working with this dataset but with different focuses than this study \citep{tundo,puccetti,delia,evans,liu}, and we made comparisons to most of these works in \citet{swift1}.
In this paper, we compare our results to the more recent \swift\ XRT Cluster Survey (SWXCS) \citep{liu}.
 In \citet{swift1}, we detected 442 extended sources in these GRB fields.  
Of these, 209 lie in the footprint of the SDSS DR8 \citep{dr8}.
We can use this SDSS data to look for over-densities in the photometric redshift distribution of the galaxies near the X-ray source.  
Using the galaxy over-density detection method described in this paper, we confirm 104 of these candidates as galaxy clusters.  Our own optical observational data, extending to deeper magnitudes, will be presented in future works.  
The structure of this paper is as follows.  Sections \ref{sec:swift} and \ref{sec:sdss} introduce the \swift\ and SDSS data, respectively.  Section \ref{sec:method} describes the method employed to detect clusters. Special cases are discussed in detail in Section \ref{sec:cases}.  
We discuss the properties of the cluster candidates with confirmed SDSS galaxy over-densities in Section \ref{sec:properties},
including the overall redshift distribution (\textsection\ \ref{subsec:overall}),
a comparison of the X-ray luminosity to the optical properties (\textsection\ \ref{subsec:xlum} -- \ref{subsec:xlumvsopt}),
and the red sequence feature common amongst galaxy clusters (\textsection\ \ref{subsec:redseq}). 
We also compare our results to the literature.
In Section \ref{subsec:match}, we match our catalog to other cluster surveys with large footprints on the sky.
We briefly discuss cases found in a previous iteration of our X-ray source selection algorithm that are not in our current catalog (\textsection\ \ref{sec:notcatalog}).
In Section \ref{sec:discussion}, we conclude with a summary and discussion of our results.  
We assume a cosmological model with $H_0 = 70$ kms$^{-1}$Mpc$^{-1}$, $\Omega_M = 0.3$, and $\Omega_\Lambda = 0.7$.

\section{\swift\ Cluster Catalog}
\label{sec:swift}

The \swift\ X-Ray Telescope (XRT) observations of GRB fields provide a serendipitous X-ray survey. 
In \citet{swift1}, we describe in detail the data reduction and methods used in producing the X-ray cluster catalog. 
Here, we briefly describe the reduction process and source extraction procedures.  
The \swift\ XRT has a relatively large field of view (23.4$\times$23.4 arcmin$^2$) and is sensitive in the energy range of 0.2--10 keV.
These GRB observations are of medium-depth, and randomly distributed on the sky for a total area of $\sim 120$ deg$^2$ with a median flux limit of 4$\times$10$^{-15}$ \flux.
These data are ideal for finding galaxy clusters.

We downloaded all XRT ``GRB'' observations\footnote{from the HEASARC website: \texttt{http://heasarc.gsfc.nasa.gov}.} before 2013-07-27, 
and reprocessed the data as described in \citet{swift1}. 
From this, we made images and corresponding exposure maps in different energy ranges: total (0.2--10 keV), soft (0.5--2 keV), and hard (2--10 keV).
Sources were detected in the images using the CIAO tool \texttt{wavdetect} using a significance of detection threshold of 10$^{-6}$. 
We excluded GRBs by matching the known GRB positions.  
To distinguish clusters from AGNs, we modeled the surface brightness profiles and compared to $\beta$-models representing a range of cluster masses and redshifts. 
We defined an extended source to be a cluster candidate if it had a S/N ratio $\geq 4$, a minimum net photon count of 20, and a size that is at least 3$\sigma$ above the mean size of corresponding point sources at the same off axis angle.  
Since our cluster detection criteria require a minimum of 20 photons, all the cluster candidates should be real astrophysical sources. False positives should arise only from confusing point AGN with extended sources.
The catalog of these 442 cluster candidates is given in Table~4 of \citet{swift1}.

\section{SDSS Data}
\label{sec:sdss}
SDSS DR8 provides the optical data for this study.  
The publicly available data covers 14,555 deg$^2$ and has magnitude limits of $m_{lim} <$ 22.0, 22.2, 22.2, 21.3 and 20.5 mag for the $u$, $g$, $r$, $i$, and $z$
bands respectively \citep{dr5, dr8}.  
Of the 442 cluster candidates, 209 are nominally in the SDSS DR8 footprint. However, six candidates lie too close to the survey edges and were not considered further (see Figure \ref{fig:incsource}), to leave us with 203 candidates in the sample.
We downloaded catalogs of the positions, magnitudes, photometric redshifts, and available spectroscopic redshifts for all galaxies within a 40\arcmin\ radius of the cluster candidate centers to provide both source and background regions.  A small fraction of these regions are incomplete in coverage due to either being on the edge of the SDSS sky coverage, bright object masks or cosmic ray masks (we discuss these in detail in Section \ref{subsec:incomplete}).

\section{Method}
\label{sec:method}

We searched for galaxy over-densities in three-dimensional space using the galaxy positions and photometric redshifts provided by SDSS DR8. 
Where available, spectroscopic redshifts are used in place of the photometric redshifts. 
For each cluster candidate, we considered source regions with radii of 2\arcmin\ and 3\arcmin\ centered on the X-ray centroid. 
We then selected the galaxies within the prescribed source regions and separated them into redshift bins of width $\Delta z = 0.05$. 
The SDSS photometric redshifts include uncertainties, which have an average of $\sim0.12$ \citep{dr8}, larger than this binsize; however, the accuracy of the cluster redshift is improved by $\sqrt{N_{Net}}$ and is typically smaller than $\Delta z = 0.05$, where $N_{Net}$ is the number of galaxies above the background in the 2\arcmin\ or 3\arcmin\ source region and within the redshift bin. 
We search for over-densities by comparing this redshift distribution to that of the local background. 
We measured the local background in an annulus extending from 25\arcmin\ to 40\arcmin\ and centered on the \swift\ sources, where the inner radius was chosen to be sufficiently distant from the cluster regions to avoid contamination from the cluster. 
Local backgrounds are needed in our analysis due to cosmic variance and any non-uniformity in the exposure depth. 
 An example of this variability can be seen in Figure \ref{fig:incsource} where there is a distinct difference in the galaxy density between SWCL J215423.1 (top left) and SWCL J015021.7 (top right). 
From the background annulus of each source, we chose 100 random regions equal in size to the source region (so with radii of 2\arcmin\ or 3\arcmin) and calculated the mean and standard deviation of the galaxy count in each redshift bin. 
This Monte-Carlo approach is more general because it does not assume background distributions simply described by Poisson fluctuations.

We compare the average background count and its standard deviation per redshift bin to the redshift distribution of the source region.
We consider an over-density peak significant if the number of galaxies in a redshift bin is 3$\sigma$ above the averaged background and if the galaxy count is at least seven. 
Setting a minimum galaxy count is important to reduce the false positives for higher redshift clusters ($z \gtrsim 0.6$), where the number of detected galaxies is low so that the standard deviation is rather small, and the 3$\sigma$ cut can be affected more by systematic uncertainties, such as the incompleteness of galaxies in SDSS for higher redshifts.
For potential clusters with multiple over-density peaks, we select the most significant redshift bin for that candidate. 
If the original distribution did not pick up a significant over-density peak, we shift all of the distribution bins to the left by 0.025 and search for over-density peaks as before.
 Using this method, we search at most 40 redshift bins of width $\Delta z = 0.05$ for each cluster candidate.  Since all of our optically confirmed clusters are below $z=0.8$, we essentially have at most 32 redshift bins for detection.  For the 203 cluster candidates, the expected number of $3\sigma$ false positives is $32*203*0.0015 \lesssim 9$ or approximately 9\% of the identifications.
The cluster redshift is determined by averaging the galaxy redshifts in the over-density bin. 
However, this includes foreground and background galaxies in the bin so the cluster redshifts determined favor the center of the over-density bin. 
To measure a more accurate cluster redshift, we shift the center of the bin $\pm 0.025$ using increments of 0.001. 
For each case, we perform the same method as before. 
We keep the most significant over-density as the best redshift estimate and report cluster details in Table \ref{tab:confz}. 
This maximizes the likelihood of our method measuring accurate redshifts while maintaining a low number of tests to minimize the number of false positives.

\begin{figure}
    \centering
     \includegraphics[width=0.8\textwidth]{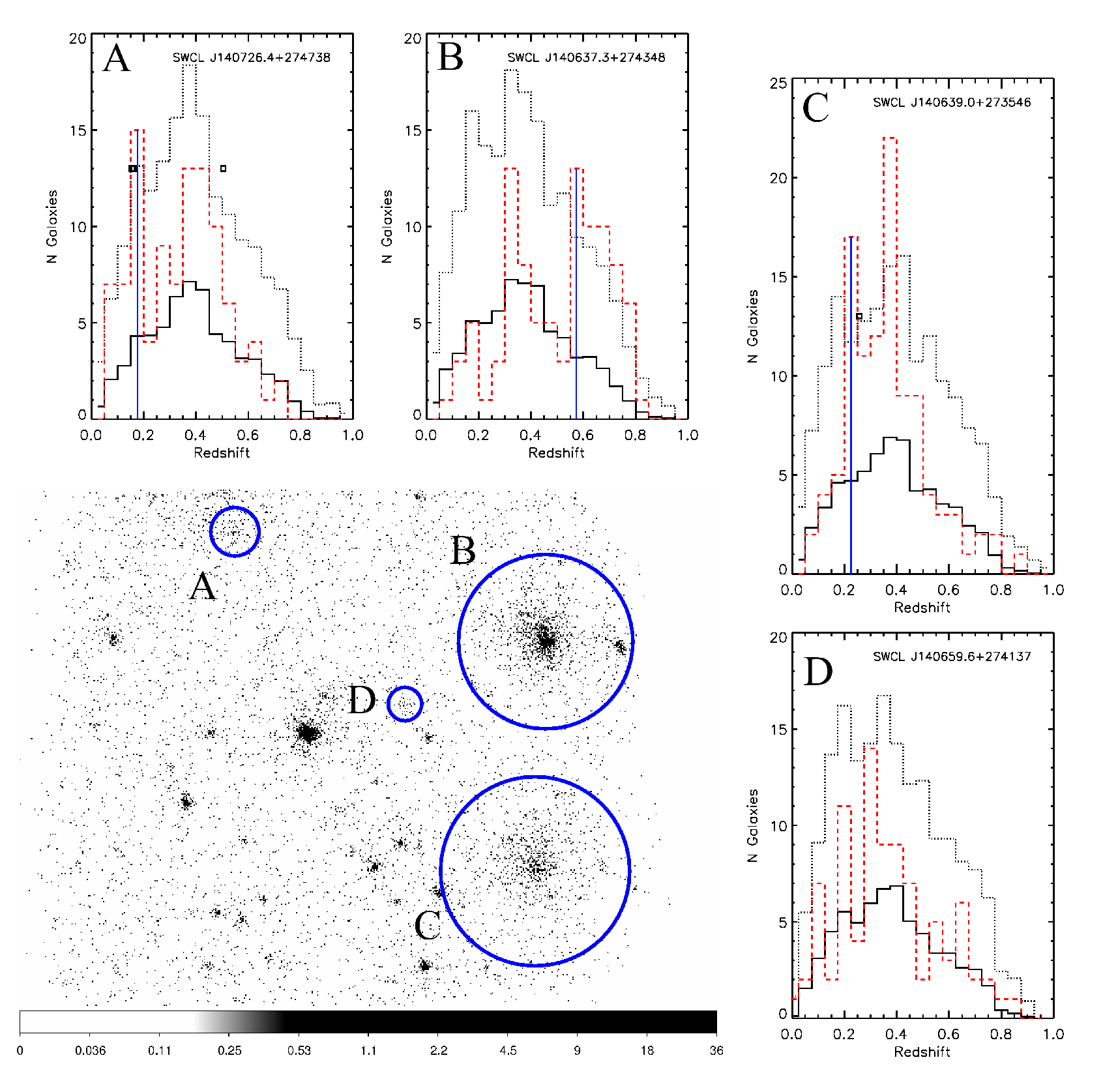}
\caption{\label{fig:sw1} SDSS identification of \swift\ clusters in the GRB060204b field. (bottom left) The X-ray image with extended sources A$-$D indicated by the blue circles.   (other panels) Photometric redshift distributions (red dashed line) of galaxies within 2\arcmin\ of source center are shown for the four labeled sources.  Also shown are the averaged distribution of random positions of field galaxies (solid black line) along with 3$\sigma$ above the average (dotted black line).  The black squares indicate any SDSS spectoscopic redshifts for galaxies within the 2\arcmin\ region.  The blue vertical lines indicate the confirmed redshifts. In this field, A is confirmed at $z$ = 0.174 at a significance of 4.46$\sigma$, B is confirmed at $z$ = 0.600  (5.34$\sigma$), C is confirmed at $z$ = 0.232 (5.95$\sigma$), and D has no SDSS confirmation.}
\end{figure}

\begin{figure}
    \centering
    \includegraphics[width=1.0\textwidth]{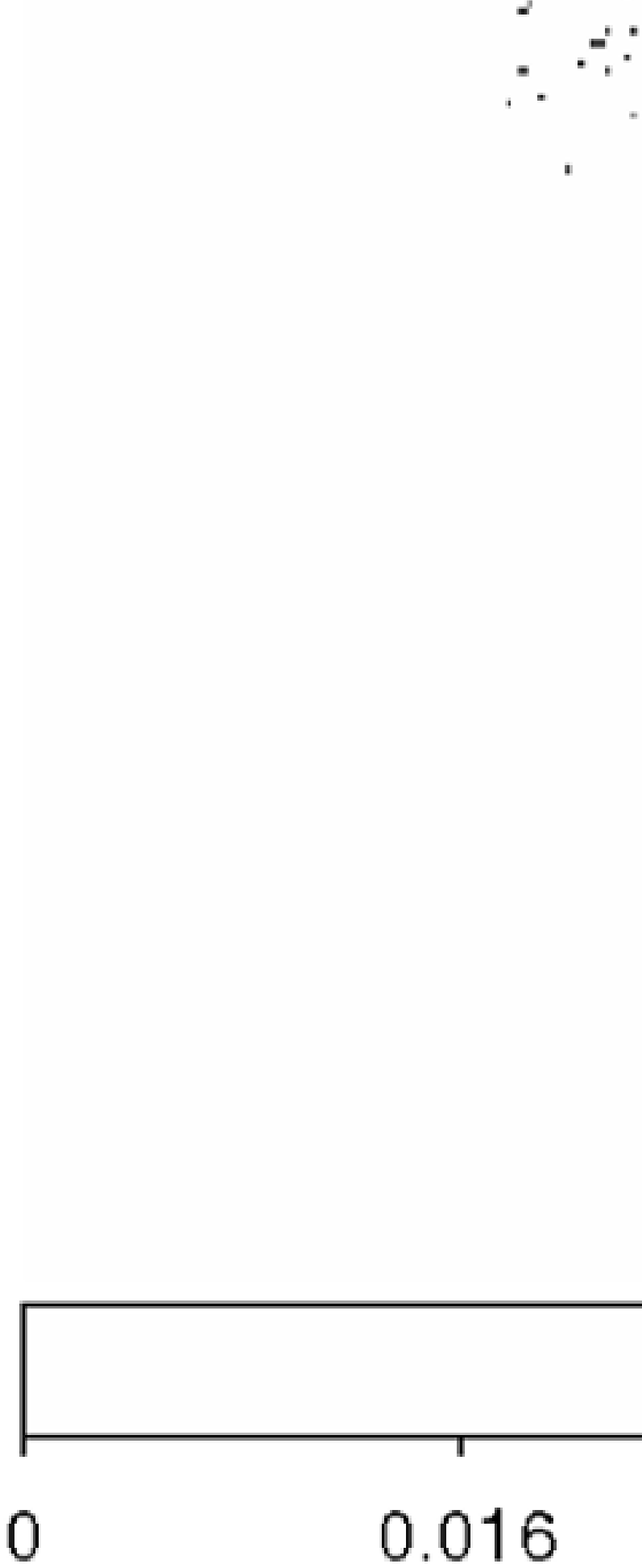}
\caption{\label{fig:sw2} SDSS identification of \swift\ clusters in the GRB061110a field. (bottom left) The X-ray image with extended sources A$-$E indicated by the blue circles.   (other panels) Photometric redshift distributions (red dashed line) of galaxies within 2\arcmin\ (A,D,E) and within 3\arcmin\ (B,C) of source center are shown for the five labeled sources.  Also shown are the averaged distribution of random positions of field galaxies (solid black line) along with 3$\sigma$ above the average (dotted black line).  The blue vertical lines indicate the confirmed redshifts. In this field, A is confirmed at $z$ = 0.467 at a signifcance of 3.56$\sigma$, B is confirmed at $z$ = 0.292 (5.16$\sigma$),  C is confirmed at $z$ = 0.162 (4.00$\sigma$), D is confirmed at $z$ = 0.714 (3.84$\sigma$), and E is confirmed at $z$ = 0.499 (6.44$\sigma$).}
\end{figure}

Our results are listed in Table \ref{tab:confz}, where '...' indicates cases where no 3$\sigma$ peak was detected for either the 2\arcmin\ or 3\arcmin\ source region sizes. We confirm SDSS over-density peaks for 104 of the 203 cluster candidates in the SDSS footprint.
This does not mean that the remaining 105 are not clusters because the SDSS data are only deep enough to confirm $z \lesssim 0.4$ clusters consistently.
Examples of our SDSS cluster confirmation for two \swift\ XRT images of GRB060204b and GRB061110a are shown in Figures \ref{fig:sw1} and \ref{fig:sw2}, respectively. 
The blue circles mark the positions of the extended X-ray sources, and the galaxy redshift distributions for the \swift\ cluster candidates (red dashed lines) are shown in separate panels together with the local mean redshift distribution and the $3\sigma$ level above the local mean.

\begin{figure}[h!]
    \centering
    \includegraphics[width=0.5\textwidth]{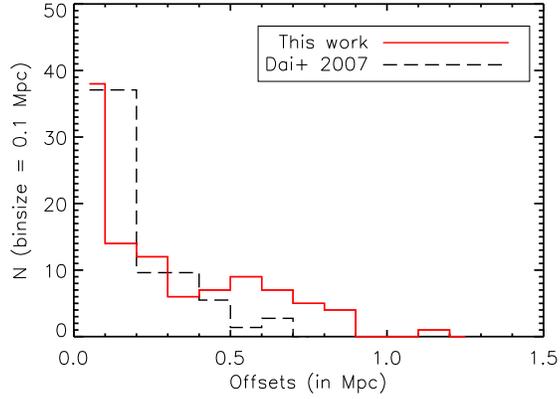}
\caption{\label{fig:bcg} Offsets between BCGs and the X-ray source centers for the 104 confirmed clusters (red, solid line). For comparison, the black, dashed line is the result from \citet{dai07}, normalized to our study.}
\end{figure}

\begin{figure}[h!]
    \centering
    \includegraphics[width=0.7\textwidth]{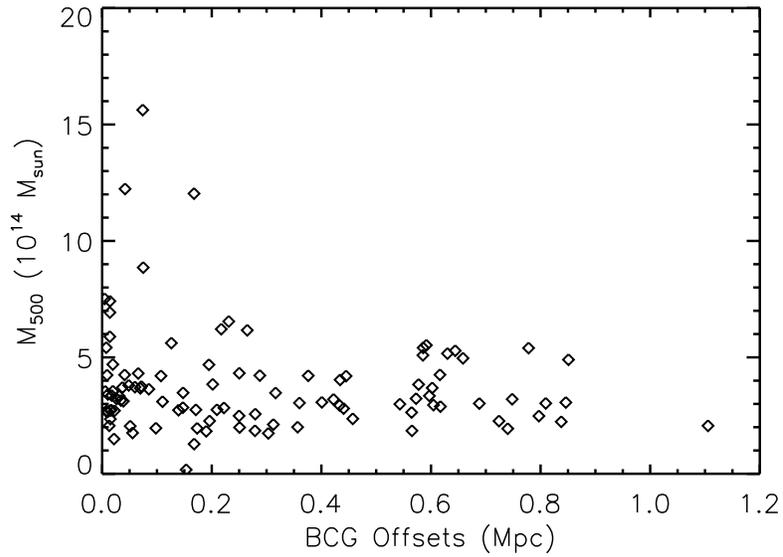}
\caption{\label{fig:massvsbcg} Mass versus BCG offsets for the 104 confirmed clusters.}
\end{figure}

For each candidate with a confirmed over-density peak, we measure the distance between the BCG and the X-ray source center, and calculate the distribution of the physical offsets (in Mpc) between the BCG and the X-ray source center (Figure \ref{fig:bcg}).  
We select the BCG as the galaxy with the brightest SDSS $r-$band magnitude in the source region and the $\Delta z = 0.05$ redshift bin.  
In Table \ref{tab:confz}, we list the SDSS $r-$band apparent and absolute magnitudes of the BCGs and their physical offsets to their corresponding X-ray centers. 
It is often assumed that both BCGs and X-ray centroids define the center of the cluster potential wells and thus, their locations should overlap. However, in studies of optically selected galaxy groups it has been found that the brightest halo galaxies can be satellite galaxies instead of central galaxies (e.g., \citealt{skibba,weinmann,bosch,pasquali}). \citet{skibba} also found that the number of brightest halo galaxies that were not central increased when studying a higher group mass bin. We show our calculated mass (details in Section \ref{subsec:xlum}) versus BCG-to-X-ray offset in Figure  \ref{fig:massvsbcg} and find the opposite to be true for our sample, that higher mass clusters have low BCG offsets.
The BCG-to-X-ray centroid distance distribution exhibits a broad tail extending from a compact core (e.g., \citealt{bcg2, bcg1, dai07}).  
We compare our offset distribution with that of 2MASS clusters \citep{dai07}, which is normalized to match our sample count, 
and find that the distributions are similar.  

Of the 104 confirmed clusters, there are 73 cases where our method found SDSS galaxy over-densities for both the 2\arcmin\ and 3\arcmin\ regions. 70\% of these cases agree within the redshift range $\delta z \lesssim 0.05$, where we assigned the more significant peak as the redshift of the cluster. The twenty-four cases where the 2\arcmin\ and 3\arcmin\ regions both detect an over-density peak but the peaks are located in different redshift bins are listed in Table \ref{tab:cases}. 
We compared both the significance of the over-density peak and the BCG-to-X-ray center offset (in Mpc) for both possibilities and assign a score as $(\sigma_2 - \sigma_3)/3 - (\rm{Offset}_2 - \rm{Offset}_3)/$Mpc for each cluster in Table~\ref{tab:cases}. 
If the score is positive (negative) then we assign the redshift to that  measured by the 2\arcmin\ (3\arcmin) region.  
The final assigned redshifts are shown in bold in Table \ref{tab:cases}. 

\section{Special Cases}
\label{sec:cases}
In this section, we discuss the situation where multiple \swift\ cluster candidates are closely located in angular position and redshift space and check for complications in the SDSS confirmation method.
We also discuss how we treat clusters with incomplete SDSS coverage of either the source or background areas. 

\subsection{Close Pairs of \swift\ Cluster Candidates}
\label{subsec:close}
\begin{figure}
    \centering
    \includegraphics[width=1.0\textwidth]{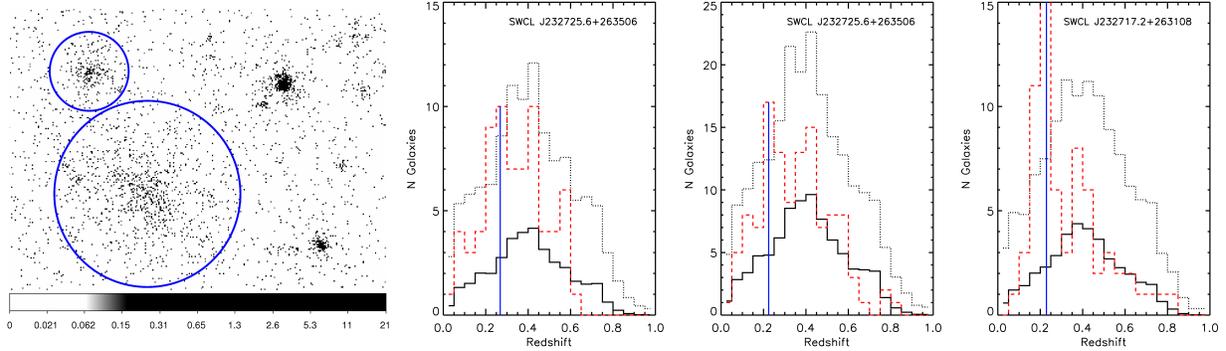}
\caption{\label{fig:nearbycont} An example of nearby contamination.  Left: the \swift\ X-ray image. SWCL J232725.6 (upper source, left middle: 2\arcmin\ detection, right middle: 3\arcmin\ detection) and SWCL J232717.2 (lower source, right panel, 3\arcmin\ detection) have centers that are 4.38\arcmin\ apart. Both source size regions for SWCL J232717.2 detected a galaxy cluster at $z \sim 0.223$ and the 3\arcmin\ source size region for SWCL J232725.6 detected a cluster at $z \sim 0.227$. SWCL J232725.6 also detected a galaxy cluster at 0.257 via 2\arcmin\ detection. We determine the 3\arcmin\ confirmation for SWCL J232725.6 is contamination and that SWCL J232725.6 is a cluster at $z \sim 0.257$.
}
\end{figure}

\begin{figure}
    \centering
    \includegraphics[width=0.8\textwidth]{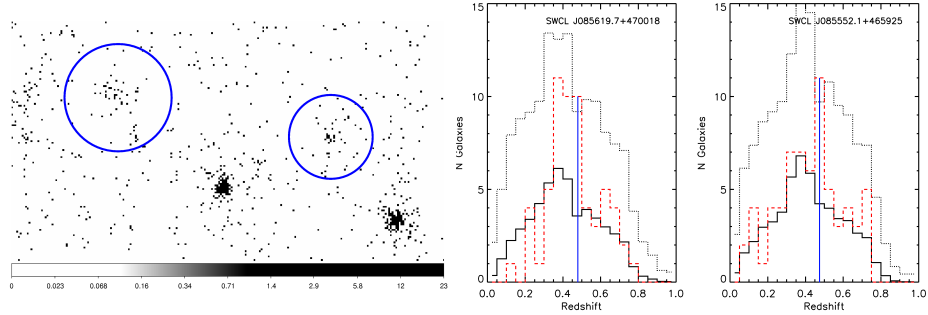}
\caption{\label{fig:nearbycont2} An example of nearby contamination. Left: the \swift\ X-ray image. SWCL J085619.7 (middle panel, left source) and SWCL J085552.1 (right panel, right source) have centers that are 4.8\arcmin\ apart. The 2\arcmin\ source size region for SWCL J085552.1 and the 2\arcmin\ and 3\arcmin\ source size regions for SWCL J085619.7 all have significant over-density peaks at $z \sim$  0.48. The distributions from the 2\arcmin\ regions are shown here.  We conclude that SWCL J085619.7 and SWCL J085552.1 are both clusters. }
\end{figure}

We found four cases with \swift\ cluster candidate pairs within 6\arcmin\ of each other and with similar redshifts ($\delta z\lesssim0.05$). 
With the possibility of overlapping source regions, these could result in false positive detections or redshift mis-assignments and thus require a closer examination.  
For each case, we compare the following for the cluster properties in question: the significance of the over-density peak, confirmation using the other source region size (so either 2\arcmin\ or 3\arcmin), the BCG$-$to$-$X-ray offset, any other redshift peaks over $3\sigma$ (i.e.\ multiple peaks), and both X-ray images and galaxy positions. 
 We list the conclusions of our analysis in Table~\ref{tab:close}, and show two examples in 
Figures \ref{fig:nearbycont} and \ref{fig:nearbycont2},  where we also include a discussion of our conclusions in the figure captions. 

SWCL J215423.1 and SWCL J215413.2 are separated by a distance of 2\marc7 and both have a high significance peak ($> 6\sigma$) around $z \sim 0.17$, and SWCL J215413.2 has a second galaxy over-density peak at $z=0.227$ (with significance $\sigma = 4.60$).  It is likely that the $z \sim 0.17$ peak for SWCL J215413.2 is contaminated by the cluster members of SWCL J215423.1.
We assign $z=0.169$ for SWCL J215423.1 and $z=0.227$ for SWCL J215413.2.

SWCL J232717.2 and SWCL J232725.6 are separated by a distance of 4\marc4 and both have a high significance peak ($> 6\sigma$) at $z \sim 0.22$, and SWCL J232725.6 has a second galaxy over-density peak at $z=0.257$ (with significance $\sigma = 5.71$).  It is likely that the $z \sim 0.22$ peak for SWCL J232725.6 is contaminated by the cluster members of SWCL J232717.2.  We assign $z=0.223$ for SWCL J232717.2 and $z=0.257$ for SWCL J232725.6 (Figure \ref{fig:nearbycont}).

 In the remaining two cases, there are no additional over-density peaks in either of the close pairs.  
For the SWCL J092730.1 and SWCL J092719.6 pair, we detected over density peaks at $z \sim 0.396$ ($z \sim 0.331$) using 2\arcmin\ (3\arcmin) source size regions for SWCL J092719.6, but only a peak at $z \sim 0.303$ using 2\arcmin\ region for SWCL J092730.1.  Further examining the BCG offsets, significance of the peaks and the X-ray images, we determine that both are clusters, possibly merging. We reach a similar conclusion for the SWCL J085552.1 and SWCL J085619.7 pair (Figure \ref{fig:nearbycont2}).

\subsection{Incomplete SDSS Coverage}
\label{subsec:incomplete}

If a cluster candidate is close to the edge of the SDSS footprint, then the 40\arcmin\ region surrounding the X-ray source's center will appear to be artificially devoid of galaxies.  In addition, there are masking regions associated with bright sources and cosmic rays in the images.  
To check for coverage completeness for each candidate cluster, the galaxy distribution within 40\arcmin\ of X-ray source center was visually inspected.

\begin{figure}
    \includegraphics[width=0.9\textwidth]{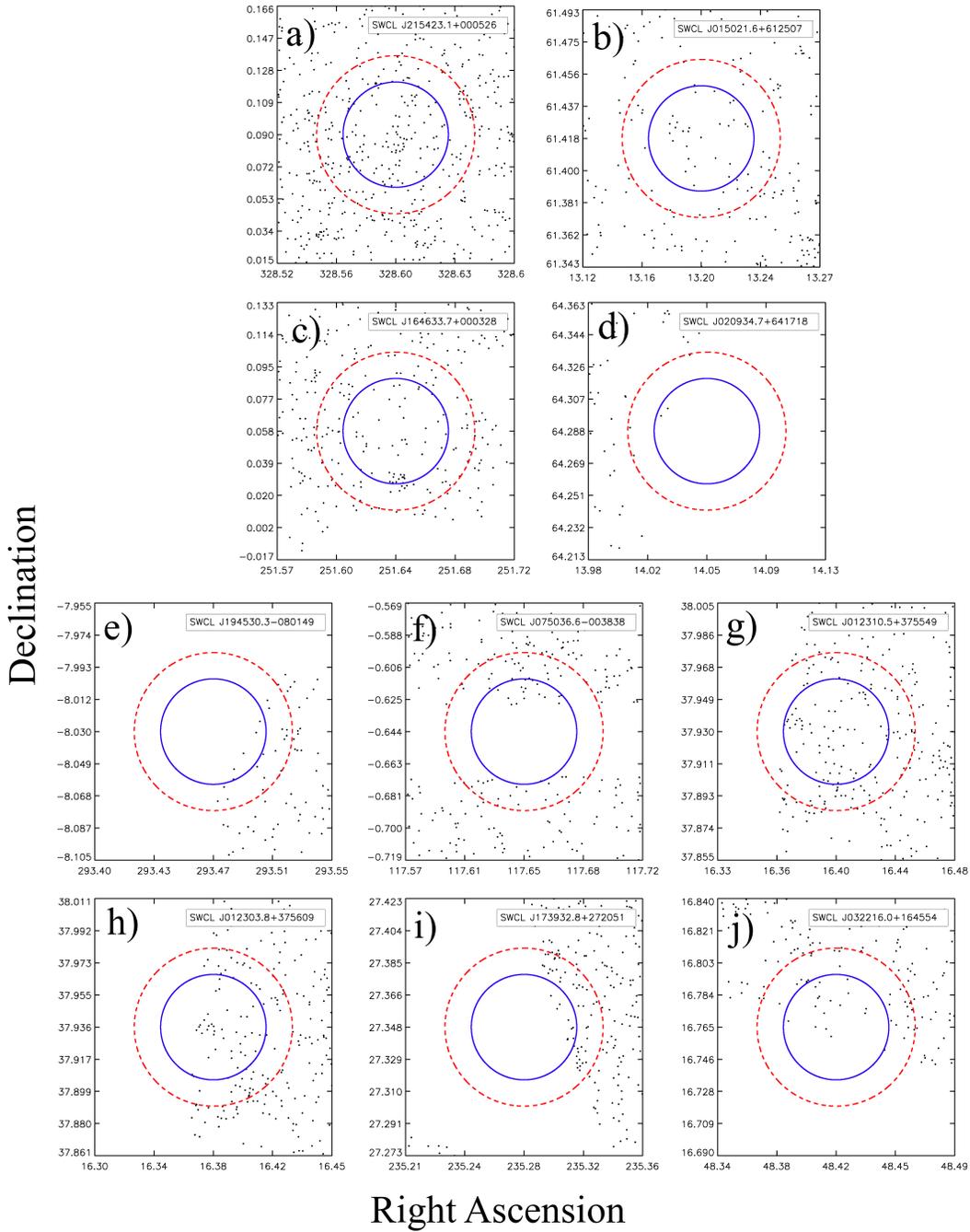}
\caption{\label{fig:incsource} Distributions of SDSS galaxies relative to source regions.  Each dot is a galaxy and the source regions are indicated by the red, dashed lines (3\arcmin\ radius) and blue, solid lines (2\arcmin\ radius). The top row (a and b) shows examples of cluster candidates with complete source regions in SDSS DR8.  The remaining cases show various degrees of incompleteness: c and g are complete for the 2\arcmin\ region and thus are included in the survey. The remaining six are too incomplete and are rejected from the survey, so that the number of cluster candidates in this survey is 203.}
\end{figure}

Eight of the candidates have incomplete 3\arcmin\ source regions as shown in Figure \ref{fig:incsource}.  Each dot represents a galaxy, the blue solid circles represent the 2\arcmin\ regions and the red dashed circles represent the 3\arcmin\ regions, indicating the largest source region size considered in our analysis.  
Two of the eight cases (c and g of Figure \ref{fig:incsource}) have complete 2\arcmin\ source regions and thus, are mostly complete in SDSS and are included in this study. The other six incomplete source regions shown in Figure \ref{fig:incsource} are excluded from this study. We are examining these cases by performing our own follow-up observations. This lead to the exclusion of 6 candidates (SWCL J012302.8, SWCL J020934.7, SWCL J032216.0, SWCL J075036.6, SWCL J173932.6 and SWCL J194530.3), as discussed earlier in Section \ref{sec:sdss}.

\begin{figure}
    \includegraphics[width=1.0\textwidth]{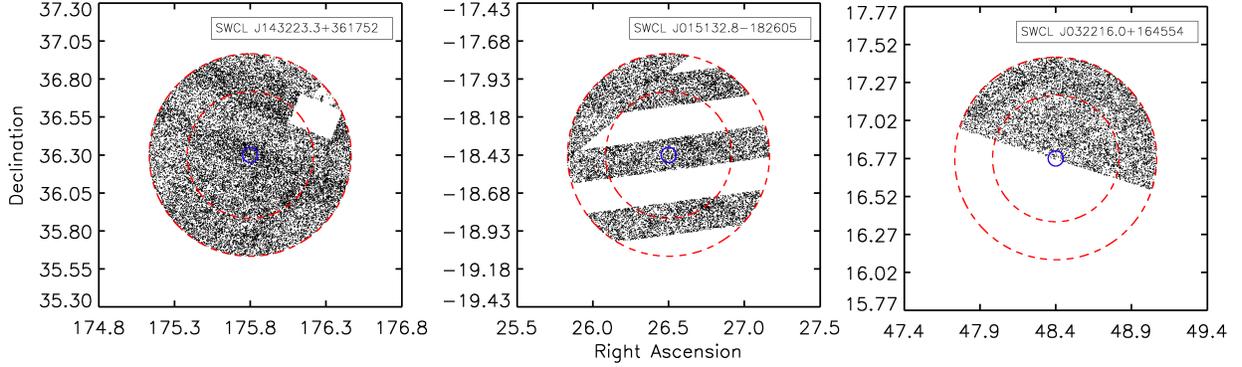}
\caption{\label{fig:incback} Incomplete backgrounds of SDSS fields.  Each dot is a galaxy, the blue, solid circle represents the 3\arcmin\ region and the red, dashed circles encompass the background annulus from 25\arcmin\ to 40\arcmin.  Left: an example of background that is $\geq 90$\% complete. Middle, Right: Examples of background regions that are $>10$\% incomplete.}

\end{figure}

There are 33 candidates with incomplete background annuli, and Figure \ref{fig:incback} shows a few examples. Other examples of incomplete backgrounds are similar in shape and incompleteness to those that are shown.  
Each dot represents a galaxy, the blue solid circles are 3\arcmin\ regions containing the source, and the red dashed circles denote the background annulus with inner and outer radii 25\arcmin\ and 40\arcmin.  
Since we consider 100 random regions to calculate the average local background and variance, the incomplete background causes the average count of galaxies to artificially decrease and the variance significantly increases. For these 33 cluster candidates, we considered 100 random regions that exclude the incomplete areas.  The redshifts and cluster details reported in Table \ref{tab:confz} reflect the results of this augmented method.

\section{Properties of the SDSS Confirmed \swift\ Clusters}
\label{sec:properties}

Here we describe various properties of the 104 SDSS confirmed clusters, match them to existing catalogs and discuss the implications for cluster science.

\subsection{Overall Redshift Distribution}
\label{subsec:overall}
\begin{figure}
    \centering
    \includegraphics[width=0.7\textwidth]{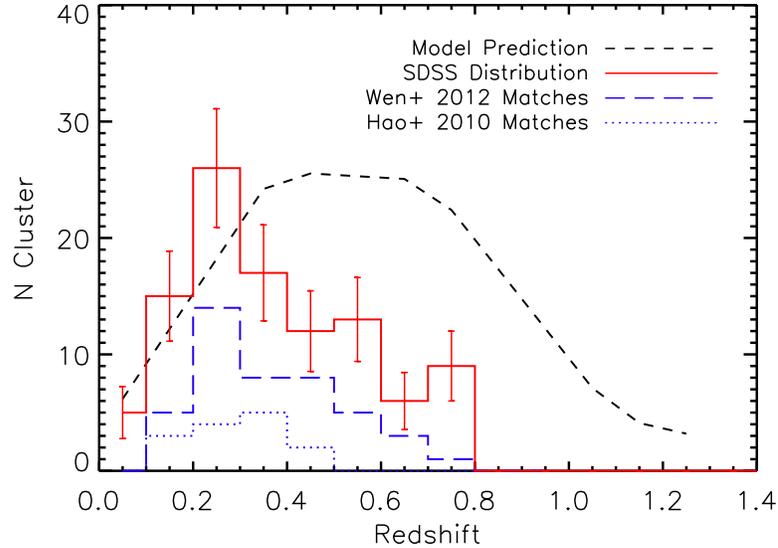}
\caption{\label{fig:tinker} Observed redshift distribution (red, solid) as compared to a model prediction (dashed, \citealt{Tinker}).  Standard error bars of $\sqrt{N}$ are used.  The dashed and dotted histograms show redshift distributions of matches between our clusters and the \citet{wen} and GMBCG \citep{gmbcg} catalogs.}
\end{figure}

Figure \ref{fig:tinker} shows the distribution
of our SDSS confirmed cluster redshifts, as compared to the \citet{Tinker} model for the mass function and redshift evolution of dark matter halos with masses ranging from $10^{14} h^{-1} M_\odot$ to $10^{15} h^{-1} M_\odot$, assuming $\Omega_m =0.25$, $\sigma_8=0.9$, $h = 0.72$, and $\Delta = 2000$.
Although the cosmological parameters are slightly different from the ones used in our paper, the model predictions are sufficient for our current qualitative comparison, providing us with a model to test the completeness of our catalog.  
Figure \ref{fig:tinker} compares our observed redshift distribution to that expected from the model.  
For lower redshifts, our survey appears complete up to $z \sim 0.3$ and is 80\% complete up to $z \sim 0.4$.  
However, the observed distribution is significantly lower for higher redshifts.  
This is presumably due to the shallowness of the SDSS DR8 catalog for galaxies at these redshifts where galaxies are fainter. For the higher-z cluster candidates, we are performing our own, deeper optical observations.  

\subsection{X-ray Bolometric Luminosities}
\label{subsec:xlum}
An advantage of having an X-ray selected catalog with optical follow-up is that we can more easily identify correlations between properties in the two bands. Here, we discuss how we estimate the X-ray luminosity and compare it to the optical properties of the 104 candidate clusters with confirmed SDSS galaxy over-densities. 

We used XSPEC \citep{xspec} to convert the \swift\ X-ray count rates to luminosities, assuming a Galactic neutral hydrogen density of 5$\times10^{20}$ cm$^{-2}$, a cluster temperature of 5 keV, and an abundance of 0.4 Solar.  Assuming a $\beta$-model with $\beta$ = 0.65, we corrected the X-ray luminosities to an aperture of radius 1.0 Mpc.  In this way the aperture would be similar to that used for the optical richness calculations. The uncertainties in $L_X$, as seen in Figure \ref{fig:xlum}, are a combination of the uncertainties of the X-ray photon counts and the 25\% systematic uncertainties from changing the assumed $\beta$ value by $\pm$0.15. 
We also estimate $M_{500}$, the mass inside of a radius $R_{500}$ at which the density of the cluster is 500 times the critical density ($\rho_c$), using the relation 
\begin{multline}
\text{ln}\ L_X = (47.392 \pm 0.085) + (1.61 \pm 0.14)\ \text{ln}\ M_{500} + (1.850 \pm 0.42)\ \text{ln}\ E(z)
\\ -  0.39\ \text{ln}\ (h/0.72) \pm (0.396 \pm 0.039)
\end{multline}

\noindent from \citet{vikhlinin}, where $E(z) \equiv H(z)/H_0$. These estimates are included in Table \ref{tab:confz}.

\subsection{Optical Properties}
\label{subsec:opticalprop}
We report observed galaxy counts ($N_{Net}$) and optical richness estimates ($N_{opt}$) in Table \ref{tab:confz}. The first is the excess galaxy count above the average background in the redshift bin of width $\Delta  = 0.05$. 
$N_{opt}$ is the estimated number of galaxies with magnitude brighter than $M_*$, assuming a Schechter luminosity function (e.g., \citealt{blackburne,dai09}) with a break magnitude $M_*$ = $-$21.34 mag and slope $\alpha$ = $-$1.07 based on the results of \citet{bell03} for SDSS$-r$ band.  
We assume $M_*$ evolves with redshift using the correction $\Delta M_* = Qz$ with $Q=1.2$ as described by \citet{dai09}.

We converted the limiting apparent $r-$band magnitude $m_{lim} < 22.2$ into an absolute magnitude limit for each cluster candidate with a confirmed SDSS galaxy over-density.  
In calculating the limiting absolute magnitudes, we take into account Galactic dust extinction and also apply $K$-corrections.
We use the NED\footnote{NASA/IPAC Extragalactic Database: \texttt{ned.ipac.caltech.edu}} online calculator for Galactic extinction values based on the X-ray positions of the cluster candidates.  We apply $K$-corrections using the template models of \citet{assef10} for the elliptical galaxies, the expected dominant galaxy population in clusters.
We normalize the Schechter luminosity function for each cluster candidate using the limiting absolute magnitude and the observed background-subtracted galaxy count. We then use the redshift dependent $M_*$ to estimate the optical richness, $N_{opt}$, of the confirmed galaxy clusters.  

Since $N_{Net}$ is found using apparent radii on the sky of 2\arcmin\ and 3\arcmin, we correct $N_{opt}$ to an aperture radius of 1.0 Mpc, assuming an NFW density profile \citep{nfw}.  We do this by multiplying $N_{opt}$ by 
$F(<R_{1.0})/F(<R_{obs})$, where $R_{1.0}$ is the projected radius of 1.0 Mpc, $R_{obs}$  is the aperture radius in Mpc corresponding to either 2\arcmin\ or 3\arcmin, and $F(<R)$ is the projected fraction inside radius R. 
We used the $L_X - M_{500}$ and $M_{200} - c_{200}$ relations to estimate the scale radius $r_s$ \citep{vikhlinin,m200c200,nfw,dai10,girardi}.
The uncertainties in $N_{opt}$ are dominated by the statistical uncertainties in the galaxy counts ($\sqrt{N_{Net}}$) and are plotted in Figure \ref{fig:xlum}.
Since we calculate the excess galaxy count using a redshift bin width $\delta = 0.05$, smaller than the typical photometric redshift uncertainty of $\sim0.12$ (as discussed in Section \ref{sec:method}), we apply a correction factor of 6.01 for all clusters.
At $z > 0.6$, where any SDSS galaxies must be significantly brighter than $L_*$, this procedure produces large corrections that tend to grossly overestimate $N_{opt}$.
These clusters are indicated in Figure \ref{fig:xlum} with crosses ($\times$). 
In addition to $N_{Net}$ and $N_{opt}$, we include $r$-band absolute magnitude of the BCGs. These are listed in Table \ref{tab:confz}. 

\subsection{Optical-to-X-ray Correlations}
\label{subsec:xlumvsopt}
\begin{figure}
    \centering
    \includegraphics[width=0.7\textwidth]{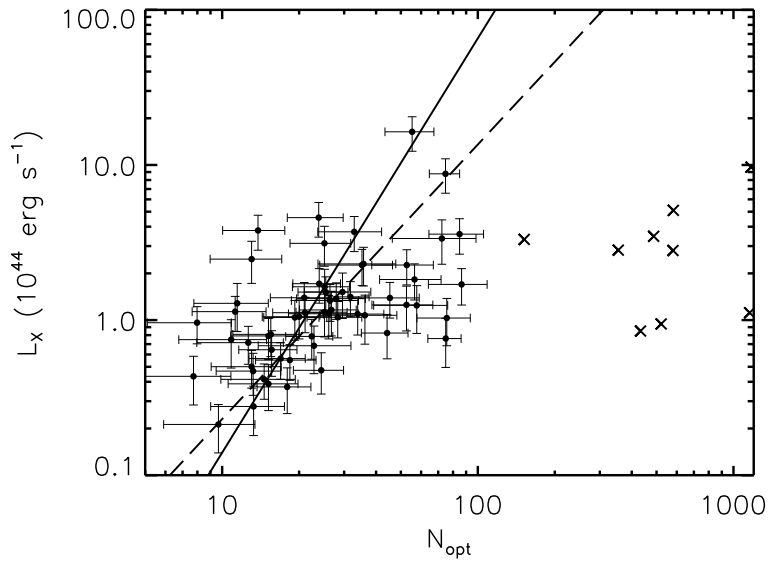}
\caption{\label{fig:xlum} X-ray Bolometric Luminosity versus Optical Richness.  The points are the confirmed $\swift$ clusters and the black lines are the best fit orthogonal regression lines for the clusters with $z < 0.6$. \textbf{The equations for the solid and dashed lines are $L_X/10^{44} = 10^{-3.53}N_{opt}^{2.67}$ and $L_X/10^{44} = 10^{-2.49}N_{opt}^{1.86}$, respectively. The former represents the best fit that includes error dependent weights and the latter represents the best fit without any weighting.} The $\times$'s indicate clusters with $z > 0.6$, most likely with over-estimated richnesses due to the magnitude limits of SDSS and the faintness of distant galaxies. For this reason, these are excluded from the linear fit. }
\end{figure}

Figure \ref{fig:xlum} shows $L_{X}$ as a function of $N_{opt}$ for cluster candidates confirmed using the 3\arcmin\ region. Both $L_X$ and $N_{opt}$ correlate well with mass, as $L_X$ probes the gas mass and $N_{opt}$ traces the galaxy members (e.g., \citealt{baxnosocs}).
\textbf{The scaling relation $L_X/10^{44} = 10^{-3.53}N_{opt}^{2.67}$ we obtain in log space using an orthogonal regression fit with error dependent weights} is shown in Figure \ref{fig:xlum} 
\textbf{(solid line). Without the error dependent weights, we obtain the scaling relation of $L_X/10^{44} = 10^{-2.49}N_{opt}^{1.86}$, also shown in Figure \ref{fig:xlum} (dashed line).} The orthogonal regression line minimizes the perpendicular distances between the data points and the line (e.g., \citealt{isobe}).
\textbf{Our slope of 2.67 is steeper} than the results of \citet{dai07} and \citet{kochanek}, who report slopes of 1.56 and 1.33, respectively. \textbf{Our slope of 1.86 agrees better with these two optically selected studies.}
\citet{baxnosocs} compares the slopes of the $L_X - N_{opt}$ from several studies in their Figure 17 and these slopes are \textbf{in general} steeper than what we measure here, ranging from 1.84 to 5.86. Comparisons are difficult, as each study can vary in fitting methods and in defining and estimating $N_{opt}$ and $L_X$. 
For example, \citet{baxnosocs} discuss two methods of linear regression solutions, the ordinary least-squares (OLS) bisector and orthogonal regression (e.g., \citealt{isobe}, where the OLS bisector is the line that bisects the OLS solution minimized in the Y direction (OLS(Y|X)) and the OLS solution minimized in the X direction (OLS(X|Y)). 
\citet{baxnosocs} found that the orthogonal regression line was better suited for this fit, especially in cases of large scatter. For ease of comparison and based on their results, we choose to use the orthogonal regression method. We find our slope is on the shallow end of those seen in \citet{baxnosocs}. Here we mention a few potential reasons for this difference. We define optical richness in Section \ref{subsec:opticalprop}, and \citet{baxnosocs} use a definition based on a physical aperture and apparent magnitude of member galaxies. Furthermore, we use a bolometric X-ray luminosity and they use $L_X$ in the [0.1$-$2.4 keV] energy band. Both \citet{dai07} and \citet{donahue} use bolometric X-rayluminosities and report slopes of 1.56 and 3.60, respectively. 
Since clusters are virialized objects, the predicted scaling relation between $L_X$ and mass is $L_X \propto M^{4/3}$, $M \propto N_{opt}$, and $L_X \propto N_{opt}^{4/3}$ (e.g., Kaiser 1991).  However, with the presence of non-gravitational effects, the scaling relations deviate from the above predictions.  One of the recent measurement between $L_X$ and total mass and stellar mass yields $L_X \propto M^{1.85}$ and $L_X \propto M_*^{2.53}$ (Anderson et al.\ 2015) and our measured slope \textbf{$L_X \propto N_{opt}^{2.67}$ agrees with the $L_X$--$M_*$ relation but deviates from the $L_X$--$M$ relation of Anderson et al.\ (2015).
And our measured slope of $L_X \propto M^{1.86}$, using a fit without error dependent weights, agrees with the $L_X$--$M$ relation.}
Since all of these measurements have their own systematic uncertainties, we do not expect an exact match in the slopes at this stage.  However, those studies that measured extreme steep slopes, 4--5, for the $L_X$--$N_{opt}$ relation were subject to more severe systematic uncertainties.

\begin{figure}
    \centering
    \includegraphics[width=0.7\textwidth]{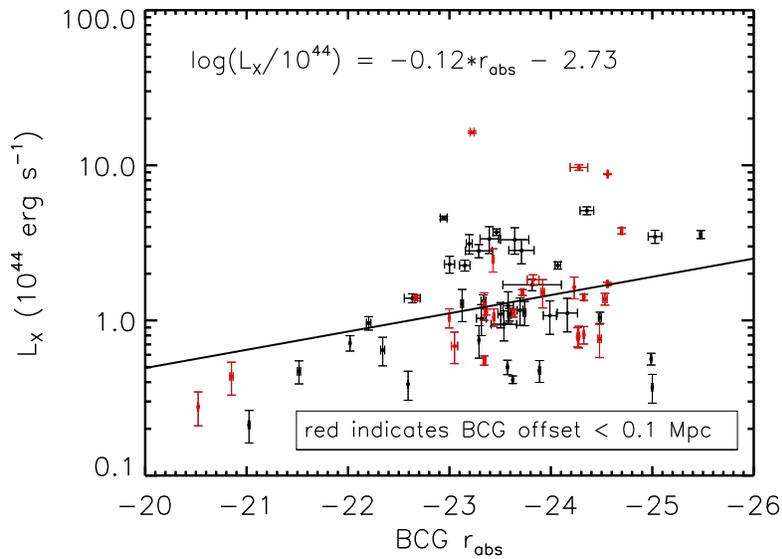}
\caption{\label{fig:xlumvsbcgmag} X-ray Bolometric Luminosity as a function of BCG Absolute Magnitude.  Here we observe a roughly positive trend, with Spearman's correlation coefficient and probability of $r_s = -0.26$ and 0.03. This indicates a positive correlation (with large spread) between gas mass (from $L_X$) and BCG luminosity (from $r_{abs}$). The red points indicate BCG-to-X-ray-center offsets of $< 0.1$ Mpc and show a higher correlation than offsets of $> 0.1$ Mpc, with $r_s = -0.42$ and $r_s = -0.17$ and probabilities of 0.04 and 0.27, respectively.}
\end{figure}

We show $L_X$ as a function of BCG $r_{abs}$ in Figure \ref{fig:xlumvsbcgmag}. 
Here we test the correlation of luminosity and mass of the ICM with the luminosity of the BCG.
Although there is a large scatter, we observe a general positive correlation when we fit the data with the orthogonal regression method (e.g., \citealt{isobe}). The data is well correlated, with a Spearman's correlation coefficient of $r_s = -0.26$ and a probability of 0.03. 
Points shown in red indicate the BCGs with smallest BCG-to-X-ray center offsets.
These clusters show less scatter than clusters with greater BCG-to-X-ray offsets, with correlation coefficients of $-0.42$ and $-0.17$ and probabilities of 0.04 and 0.27, respectively. 

\subsection{The Red Sequence}
\label{subsec:redseq}
\begin{figure}
    \centering
    \includegraphics[width=0.9\textwidth]{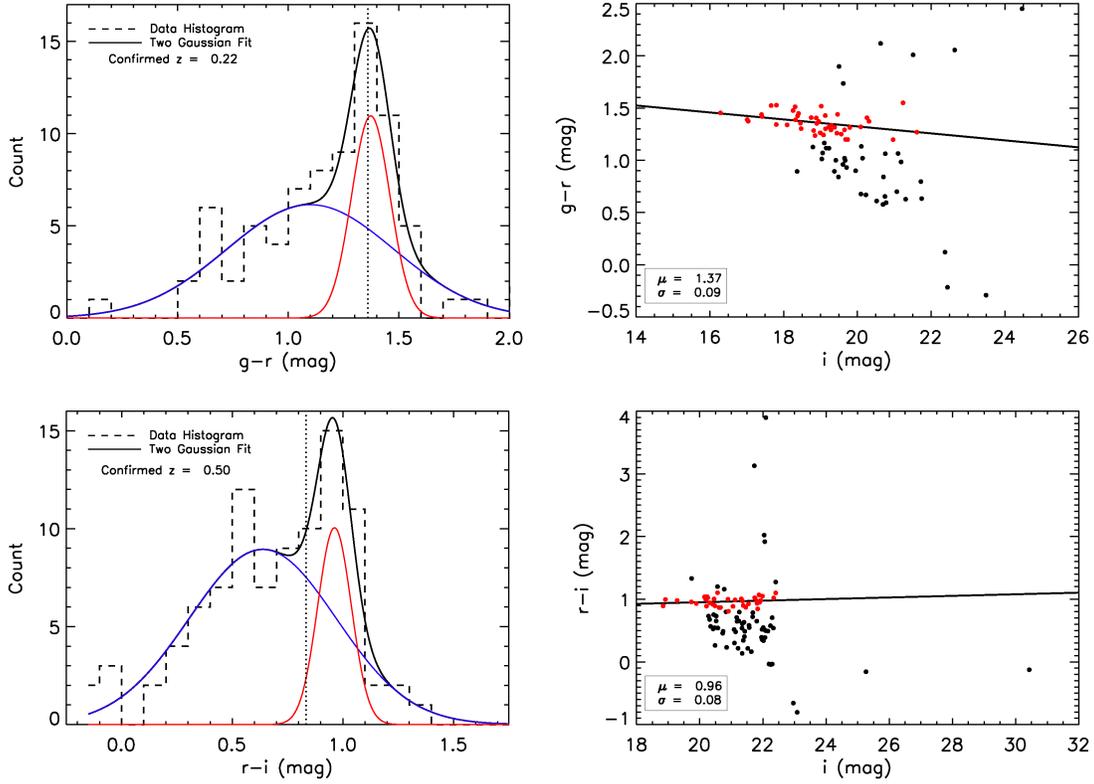}
\caption{\label{fig:colorplot} Color distributions and color magnitude diagrams.  Left: Color distributions showing clear red sequences (red Gaussian fit) and field galaxies (blue Gaussian fit), in g $-$ r for SWCL J232717.2 (top) at $z = 0.223$ and r $-$ i for SWCL J222438.0 at $z = 0.499$ (bottom).  Right: The corresponding CMDs with red points indicating galaxies within 2$\sigma$ of the red sequence Gaussian.   The line indicates the best fit to these points with mean $\mu$ and Gaussian dispersion $\sigma$.}
\end{figure}

We examined the color distributions for each candidate with a confirmed SDSS galaxy over-density to search for the red sequence feature seen in galaxy clusters.  In general, cluster members and the surrounding background form a bimodal color distribution that is well represented by a two Gaussian fit, as seen in Figure \ref{fig:colorplot}  
(e.g., \citealt{gmbcg,bell04}). 
The narrow, taller Gaussian is the red sequence feature, prominent in galaxy clusters. The distribution of background and/or foreground field galaxies can generally be approximated with a shallower, wider Gaussian, because the background galaxies are from different redshifts and the weak color bimodality is smoothed by redshifts.
Following \citet{gmbcg}, we shifted between the $g-r$, $r-i$, and $i-z$ colors to follow the shifting location of the Balmer break with redshift. The redshift ranges used for each color are listed in Table \ref{tab:colredrange}.  

\begin{figure}
    \centering
    \includegraphics[width=0.8\textwidth]{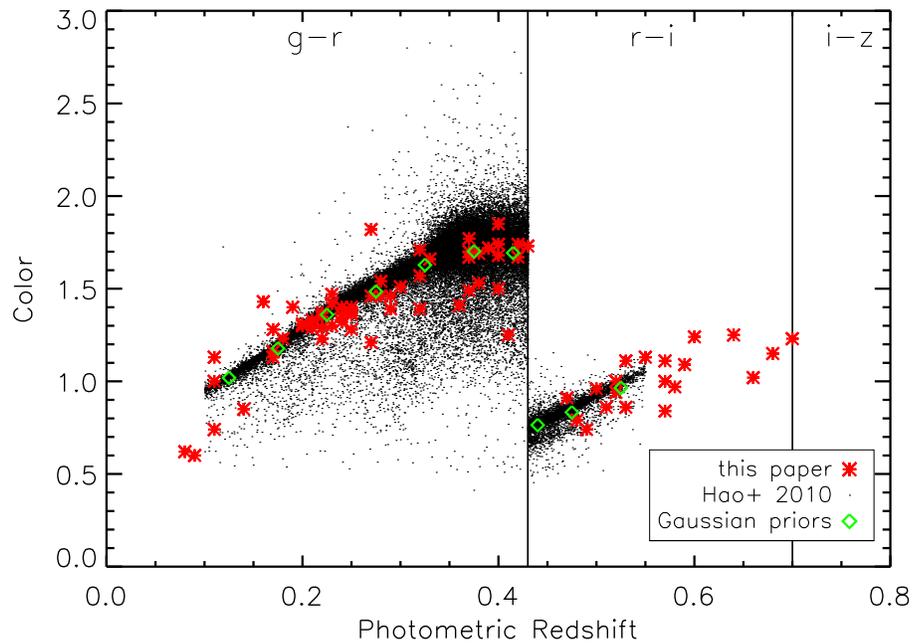}
\caption{\label{fig:meancolorplot}  Mean red sequence color as a function of redshift for every galaxy cluster (red stars) that has an SDSS galaxy over-density and convergent Gaussian fit of the red sequence. The points are clusters from the GMBCG catalog \citep{gmbcg}. The green diamonds are the color priors from the GMBCG catalog used in the two Gaussian fits. }
\end{figure}

\begin{figure}
    \centering
    \includegraphics[width=0.8\textwidth]{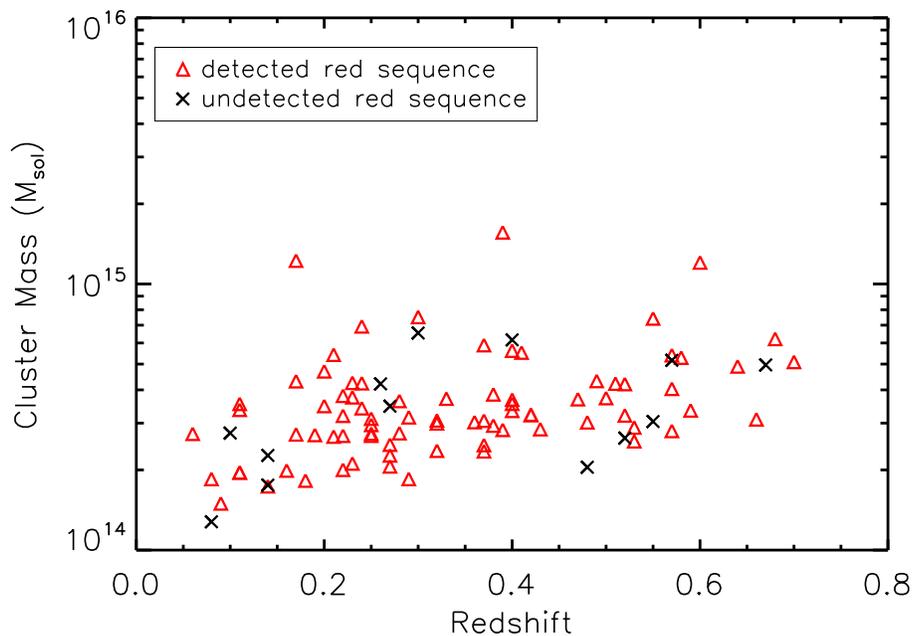}
\caption{\label{fig:massredseq} Cluster mass (in $M_\odot$) versus redshift. Here we look for a mass trend in clusters with detected red sequences in SDSS (red triangles) versus those without (black $\times$). We detect the red sequence in the 6 most massive clusters, although in general, we do not observe a mass trend. }
\end{figure}

For each cluster candidate with a confirmed SDSS galaxy over-density, we examine the color distribution using galaxies within 3\arcmin\ of the cluster center and within redshifts $z_{cl} \pm0.1$, where $z_{cl}$ is our measured cluster redshift.  
We fit the color distribution to a two Gaussian model where we allow the normalization, mean, and width of both Gaussians to vary and minimize the  $\chi^2$ statistic.
For the lower redshift clusters, $z_{cl} \lesssim 0.11$, we use a single Gaussian model with one prior for the expected color. For these cases, we do not observe color contribution from the background galaxies, likely due to the extent of the clusters on the sky.
Our two Gaussian model includes priors that constrain the width of the red sequence compared to that of the background and constrain the mean of the red sequence to favor the colors used by \citet{gmbcg} in their catalog of Gaussian Mixture Brightest Cluster Galaxy (GMBCG) clusters. We use the second prior because we know from \citet{gmbcg} how the red sequence evolves with redshift (see Figure \ref{fig:meancolorplot}).
 We calculate the second prior by averaging the color of GMBCG clusters in redshift bins of width $\Delta z = 0.05$. These are shown by green diamonds in Figure \ref{fig:meancolorplot} and by the vertical lines in Figure \ref{fig:colorplot}.
We find the red sequence feature in $\sim$85\% of the confirmed clusters. 
The cluster galaxies (red dots in Figure \ref{fig:colorplot}, right), are the galaxies with color within 2$\sigma$ of the red sequence mean. 
Two examples are shown in Figure \ref{fig:colorplot}, along with corresponding color magnitude diagrams (CMDs).  
Here we can observe the red sequence feature clearly both in the narrow histogram and in the clustering of galaxies in color-magnitude space.
We also show the mean red sequence color as a function of redshift in Figure \ref{fig:meancolorplot}.  Although there is a significant amount of scatter, Figure \ref{fig:meancolorplot} shows a general positive trend for the g $-$ r and r $-$ i colors as redshift increases, as expected.
We include GMBCG cluster colors from \citet{gmbcg} (dots) in Figure \ref{fig:meancolorplot} as well as the color priors used for our two Gaussian fit (green diamonds). In Figure \ref{fig:massredseq}, we plot the mass of clusters with confirmed redshifts, denoting clusters with a detected red sequence with red triangles and clusters without with black crosses ($\times$). We see that in general there is no difference in mass dependence. We performed a Kolmogorov-Smirnov (K-S) test to determine whether the two mass distributions are in fact different.  We find there is a 0.56 probability that the distributions are drawn from the same distribution, signifying that the mass distribution of clusters with a detected red sequence is not significantly different than those without a detected red sequence. 
However, we do detect the red sequence in the six most massive clusters, so clusters with mass greater than $6.7 \times 10^{14} M_{\odot}$.
Also, we find that the redshift distributions for detected red sequences versus undetected are significantly different, with a K-S test probability of 0.03.  

The non-detections of the red sequence in our survey has several origins. The principle problem is that many of the galaxies are relatively faint, potentially leading to large color errors compared to the width of the red sequence.
  Galaxies fainter than the magnitude limit of SDSS will not be included in the color distributions, so that the distribution may not fit the two Gaussian model.  This is especially true for more distant clusters where the observed flux is lower.
 At low redshifts, the X-ray data is also sensitive enough to include lower mass groups, which lack the well-defined red sequences of rich clusters. 
\citet{nilo} had similar issues in their study of low X-ray luminosity galaxy clusters.

\subsection{Matching with Other Catalogs}
\label{subsec:match}

We compare our catalog to other studies across multiple wavelengths to test for accuracy and completeness.  
We chose to compare to catalogs with large footprints since our \swift\ targets are scattered over a large fraction of the sky.
First, we compare with optical catalogs, starting with the 132,684 galaxy clusters in the catalog from \citet{wen}. 
Their method uses a friend-of-friend algorithm incorporating SDSS III galaxies to identify clusters and their BCGs. 
We used a matching 
angular radius of up to 1\arcmin.  
Out of 44 position matches, 41 agree in redshift within $\lvert \delta z \rvert < 0.1$ with an average redshift difference of 0.026.
These are listed in Table \ref{tab:gmbcg}.
Of the 41 redshift matches, $\sim70$\% have mass greater than the median mass of our sample, showing that in general we are matching our more massive clusters to the optically selected catalog.
\citet{wen} claimed their catalog is 95\% complete for $M_{500} > 10^{14} M_\odot$ and for the redshift range $0.05 \leq z < 0.42$, and our sample is complete for $M_{500} > 10^{14} M_\odot$ and $z <$ 0.3. The number of clusters per survey area for the overlapping conditions $M_{500} > 10^{14} M_\odot$ and 0.05 $< z <$ 0.3 are equal ($\sim 0.37$) and thus, the completeness statements for both samples are in agreement.
We also compare our confirmed cluster redshifts to the GMBCG clusters \citep{gmbcg}. Using the above matching procedure, there are 18 position matches and 15 that agree in position as well as redshift, 14 of which are in the \citet{wen} catalog. SWCL J121628.2 is not in the \citet{wen} catalog, so the right ascension, declination and separation reported in Table \ref{tab:gmbcg} refer to the GMBCG catalog. The differences in redshift and position between our catalog and these optical catalogs could arise from the fact that our clusters are X-ray selected and thus centered on the X-ray emission from the clusters as opposed to their optically selected clusters. However, some of the matches with large angular match radii ($>0$\marc$6$) might be spurious associations.

Here we compare to catalogs with selections based on X-ray, SZ and lensing data. These trace the location of the gas in the cluster (e.g., \citealt{formofgal}) and should align with our sources better than optically selected surveys, so we use a smaller matching radius of 30\arcsec.
We search for matches to X-ray selected clusters using the Meta-Catalog of X-Ray Detected Clusters of Galaxies (MCXC, 
\citealt{mcxc}).  
MCXC comprises of 1743 clusters and combines the ROSAT All Sky Survey with cluster surveys based on serendipitous observations. 
We find only two previously detected clusters, MCXC\_J1557.7+3530 at redshift $z = 0.155$ and MCXC\_J1557.7+3530 at $z = 0.360$. These match to our clusters SWCL J155743.3 at $z = 0.166$ and SWCL J025630.7 at $z = 0.374$ \citep{mcxc}. 
SWCL J025630.7 is also detected with the Atacama Cosmology Telescope (ACT) \citep{sz5} using the SZ effect.  
This is the only cluster match when compared to the 148 GHz observations by ACT of 68 galaxy clusters.  
We also compared our survey to the first release of SZ sources observed by Planck \citep{sz6} and found that SWCL J084749.4 at $z = 0.391$ is a match to PSZ1\_G213.43+31.78 at $z = 0.349$. Finally, we compared our confirmed galaxy clusters to the Deep Lensing Survey \citep{dls} but found no matches.

In \citet{swift1}, we compared our catalog of extended sources to the \swift\ XRT Cluster Survey (SWXCS) \citep{tundo}, and found 55 of 72 sources agreed in position to within 30\arcsec.
\citet{liu} expanded the analysis from using only GRB fields in \citet{tundo} to also including non-GRB fields, increasing the number of fields to $\sim3,000$ and finding 263 cluster candidates in a total solid angle of $\sim400$ deg$^2$. They cross-correlated their catalog with optical, X-ray, and SZ catalogs to match known galaxy and galaxy cluster redshifts that have similar positions. 
 Of the 442 extended sources in \citet{swift1}, 88 agree within 30\arcsec\ and 68 agree within 10\arcsec. Although the number of SWXCS cluster candidates increases from 72 to 263 \citep{tundo,liu}, we do not see a comparative increase in matches because, like \citet{tundo}, our analysis includes only \swift\ GRB fields.

We compared the positions of our SDSS confirmed galaxy clusters to their updated catalog and found 37 position matches in the SDSS footprint using a matching radius of 10\arcsec. These are shown in Table \ref{tab:liu}. In \citet{liu}, there are clusters with multiple reported redshifts that come from comparing the SWXCS to other studies. For these cases, we report the average of those redshifts in Table \ref{tab:liu}. Of those 37 matches, 23 have redshifts in both catalogs, 22 agree within $\lvert \delta z \rvert < 0.1$ and 19 agree within $\lvert \delta z \rvert < 0.05$ with an average redshift difference of $\Delta z \sim 0.033$. Furthermore, 17 of the 37 position matches were also matched to the \citet{wen} catalog and thus are also presented in Table \ref{tab:gmbcg}.

\section{Confirmed Clusters Not in Current Catalog}
\label{sec:notcatalog}
We went through various iterations of the X-ray source selection method before using the latest one discussed in \citet{swift1}.
There are 10 X-ray sources selected using a previous version of our algorithm that were confirmed as galaxy clusters using SDSS data and the method described above. 
These are listed in Table \ref{tab:notcatalog}. 
Due to slight changes in the X-ray source selection algorithm, the significances of these sources changed to be below the significance threshold and so are not included in the current catalog or in Table \ref{tab:confz}.
These clusters are worth mentioning for researchers interested in these individual clusters in future studies.

\section{Summary \& Discussion}
\label{sec:discussion}
In this paper, we present SDSS identifications (Table \ref{tab:confz}), including estimated redshifts, X-ray and optical properties, for the X-ray selected galaxy clusters we identified in \citet{swift1}.
We confirmed 104 of the 203 cluster candidates in the SDSS footprint and estimate that the catalog is 80\% complete up to $z = 0.4$.
Most of the remaining cluster candidates are expected to be at higher redshifts where the member galaxies are too faint to use SDSS for confirmations.
This sample significantly increases the number of X-ray selected clusters with confirmed redshifts and it is one of the largest uniformly-selected cluster samples, covering a total sky area of 125 deg$^2$.

We observe clear red sequences and clustering of galaxies in color-magnitude space in $\sim$85\% of the clusters.
X-ray source centers and BCG locations show good agreement, with small offsets and distributions similar to other studies.
We find clear matches with previously observed clusters. 
Our X-ray luminosities correlate well with optical properties and our $L_X - N_{opt}$ slope agrees with other estimates.
Thus, it is clear the \swift\ cluster detection technique presented in \citet{swift1} is successfully identifying extended X-ray sources that are in fact galaxy clusters.
In future studies, we will look at changing the significance threshold of the X-ray source detection \citep{swift1} as it may be too stringent currently. 
For example, we list 10 clusters in Table \ref{tab:notcatalog} that were confirmed by the SDSS data but removed from the current X-ray catalog based on the final choice for the significance threshold.

Our method combines X-ray and optical techniques and properties to confirm galaxy cluster candidates and can be used for similar studies.
In this way we find 63 new galaxy clusters in the SDSS footprint that were not detected using the optical cluster finding methods of \citet{wen} and \citet{gmbcg}. 
The next step is to look deeper in magnitude and beyond the footprint of SDSS. 
Although SDSS was a good start, the data is too shallow for higher redshift clusters and has incomplete sky coverage. There are $\sim100$ undetected \swift\ extended sources in the SDSS footprint, and still $\sim250$ outside, for a total of $\sim350$.

We have been performing our own followup observations with observing programs taking photometric data at the MDM 2.4m, Kitt Peak 4m, and CTIO (Cerro Tololo Inter-American Observatory) 4m.
We have also observed candidates with redshifts of $z \sim 0.5$ at the Magellan 6.5m and MDM 2.4m using multi-slit spectroscopic masks.
Once we derive galaxy redshifts from our data we will use the method outlined in this paper to confirm additional clusters at higher redshifts. 
Furthermore, our sample will be more statistically significant when it is more complete, which will lead to better constraints on the cluster mass function.  

\acknowledgements We acknowledge the financial support from the NASA ADAP program NNX11AD09G and NSF grant AST-1413056.

Funding for the creation and distribution of the SDSS Archive has been provided by the Alfred P. Sloan Foundation, the Participating Institutions, the National Aeronautics and Space Administration, the National Science Foundation, the US Department of Energy, the Japanese Monbukagakusho, and the Max Planck Society.

We thank the anonymous referee for the detailed report and helpful comments.

\pagestyle{empty}
\begin{landscape}
\begin{deluxetable}{rcccccccccccc}
\tabletypesize{\scriptsize}
\tablecolumns{10}
\tablewidth{0pt}
\tablecaption{The \swift\ AGN and Cluster Survey: SDSS Confirmations \label{tab:confz}}
\tablehead{
 \colhead{\swift} & \colhead{2\arcmin\ Reg.} & \colhead{3\arcmin\ Reg.} & \colhead{Conf.} & \colhead{Conf.} & \colhead{$N_{Net}$} &  \colhead{$N_{opt}$} & \colhead{BCG} & \colhead{BCG} & \colhead{Offset} & \colhead{$L_X$} & \colhead{$M_{500}$}
\\
 \colhead{Name} & \colhead{Conf. $z$} & \colhead{Conf. $z$} & \colhead{Phot. $z$} & \colhead{$\sigma$ } & \colhead{ } & \colhead{ } & \colhead{$r_{mag}$} & \colhead{$r_{abs}$} & \colhead{(Mpc)} & \colhead{(erg/s)} & \colhead{(M$_{\odot})$} 
}
\startdata

SWCL J002111.4+210438 &     0.674 &   \nodata  &    0.674 &   4.14 &   5.99 &  186.68 &  21.84 &     $-$23.24 &   0.66 &   3.52e+44 &   4.96e+14 \\
SWCL J002114.5+205943 &     0.141 &      0.074 &    0.141 &   6.08 &   6.82 &   13.19 &  16.94 &     $-$22.02 &   0.06 &   3.75e+43 &   1.75e+14 \\
SWCL J002823.6+092705 &     0.195 &      0.215 &    0.195 &   7.09 &  13.06 &   22.54 &  17.82 &     $-$23.23 &   0.02 &   7.95e+43 &   2.71e+14 \\
SWCL J003317.8+193925 &  \nodata  &      0.246 &    0.246 &   3.17 &   3.41 &    7.98 &  19.85 &     $-$22.20 &   0.61 &   9.62e+43 &   2.95e+14 \\
SWCL J005136.8+074351 &  \nodata  &      0.757 &    0.757 &   3.90 &   2.32 &  432.92 &  19.26 &     $-$26.40 &   0.74 &   8.50e+43 &   1.94e+14 \\
SWCL J012310.5+375549 &     0.749 &      0.762 &    0.762 &   4.52 &   2.37 &  521.41 &  22.29 &     $-$23.54 &   1.11 &   9.45e+43 &   2.06e+14 \\
SWCL J015751.5+170139 &  \nodata  &      0.529 &    0.529 &   4.85 &   6.20 &   52.61 &  20.66 &     $-$23.34 &   0.01 &   1.25e+44 &   2.89e+14 \\
SWCL J015752.9+165933 &     0.510 &      0.525 &    0.510 &   5.30 &  10.00 &   50.36 &  20.38 &     $-$23.15 &   0.29 &   2.26e+44 &   4.22e+14 \\
SWCL J015803.5+165005 &     0.223 &      0.233 &    0.223 &   7.46 &  11.91 &   20.55 &  16.95 &     $-$24.33 &   0.01 &   8.09e+43 &   2.69e+14 \\
SWCL J020003.8+084024 &     0.196 &   \nodata  &    0.196 &   4.10 &   6.96 &   12.48 &  16.24 &     $-$24.90 &   0.02 &   1.93e+44 &   4.70e+14 \\
SWCL J020006.4+084454 &     0.419 &   \nodata  &    0.419 &   4.01 &   8.73 &   26.16 &  19.07 &     $-$23.85 &   0.03 &   1.32e+44 &   3.21e+14 \\
SWCL J025547.9+000902 &     0.573 &      0.570 &    0.573 &   4.00 &   7.71 &   78.20 &  20.89 &     $-$23.39 &   0.63 &   3.36e+44 &   5.17e+14 \\
SWCL J025630.7+000601 &     0.374 &      0.391 &    0.374 &  15.49 &  45.33 &  112.34 &  18.22 &     $-$24.48 &   0.01 &   7.60e+43 &   2.35e+14 \\
SWCL J035259.4$-$004338 &     0.413 &      0.301 &    0.301 &   3.77 &   7.32 &   23.86 &  18.59 &     $-$22.94 &   0.36 &   4.58e+44 &   7.51e+14 \\
SWCL J054716.7+641156 &     0.271 &      0.369 &    0.369 &   6.19 &  10.37 &   44.17 &  19.36 &     $-$26.24 &   0.80 &   8.25e+43 &   2.48e+14 \\
SWCL J075043.2+310400 &     0.378 &      0.375 &    0.378 &   4.69 &  14.16 &   33.79 &  19.16 &     $-$23.59 &   0.43 &   1.09e+44 &   2.94e+14 \\
SWCL J075900.8+324449 &     0.579 &      0.751 &    0.579 &   3.70 &   7.53 &   75.21 &  21.99 &     $-$22.36 &   0.64 &   3.50e+44 &   5.28e+14 \\
SWCL J075908.5+324237 &     0.722 &   \nodata  &    0.722 &   3.99 &   5.85 &  481.81 &  21.15 &     $-$24.13 &   0.58 &   2.45e+44 &   3.83e+14 \\
SWCL J080022.5$-$085931 &  \nodata  &      0.273 &    0.273 &   3.98 &   4.25 &   11.45 &  18.94 &     $-$23.12 &   0.32 &   1.28e+44 &   3.47e+14 \\
SWCL J082113.9+320018 &     0.746 &      0.728 &    0.746 &   7.58 &   9.09 &  859.00 &  21.16 &     $-$24.28 &   0.08 &   9.68e+44 &   8.86e+14 \\
SWCL J083340.9+331118 &     0.682 &      0.675 &    0.682 &   8.42 &  18.06 &  673.84 &  20.66 &     $-$24.35 &   0.22 &   5.10e+44 &   6.21e+14 \\
SWCL J084749.4+133142 &     0.391 &      0.391 &    0.391 &   7.40 &  21.23 &   55.50 &  19.41 &     $-$23.22 &   0.07 &   1.64e+45 &   1.56e+15 \\
SWCL J084959.1+521711 &     0.201 &   \nodata  &    0.201 &   3.95 &   8.00 &   13.75 &  18.95 &     $-$22.06 &   0.15 &   1.20e+44 &   3.48e+14 \\
SWCL J085523.8+110202 &     0.317 &   \nodata  &    0.317 &   4.50 &  10.88 &   20.82 &  19.67 &     $-$22.04 &   0.40 &   1.10e+44 &   3.07e+14 \\
SWCL J085552.1+465925 &     0.477 &   \nodata  &    0.477 &   4.01 &   7.77 &   29.24 &  19.67 &     $-$23.81 &   0.69 &   1.27e+44 &   3.02e+14 \\
SWCL J085619.7+470018 &     0.479 &      0.474 &    0.479 &   3.53 &   6.41 &   23.93 &  20.27 &     $-$23.05 &   0.05 &   6.83e+43 &   2.05e+14 \\
SWCL J090714.8+351020 &     0.218 &      0.212 &    0.218 &   5.08 &  10.73 &   17.49 &  17.18 &     $-$23.57 &   0.36 &   5.01e+43 &   2.00e+14 \\
SWCL J092607.0+314854 &     0.268 &   \nodata  &    0.268 &   3.44 &   8.54 &   14.40 &  18.95 &     $-$22.61 &   0.25 &   7.47e+43 &   2.49e+14 \\
SWCL J092649.8+301346 &     0.570 &      0.567 &    0.567 &   5.33 &   7.64 &   84.84 &  18.86 &     $-$25.48 &   0.78 &   3.58e+44 &   5.40e+14 \\
SWCL J092719.6+301348 &     0.396 &      0.331 &    0.331 &   4.25 &   9.07 &   25.43 &  18.31 &     $-$23.72 &   0.04 &   1.52e+44 &   3.71e+14 \\
SWCL J092730.1+301046 &     0.303 &   \nodata  &    0.303 &   4.41 &  10.72 &   20.16 &  19.05 &     $-$23.22 &   0.23 &   3.67e+44 &   6.54e+14 \\
SWCL J092852.3+002137 &     0.535 &      0.091 &    0.091 &   4.09 &   4.32 &   13.27 &  16.91 &     $-$20.52 &   0.02 &   2.77e+43 &   1.50e+14 \\
SWCL J093041.3+170400 &     0.320 &   \nodata  &    0.320 &   4.88 &  10.96 &   20.56 &  20.00 &     $-$22.23 &   0.46 &   7.25e+43 &   2.36e+14 \\
SWCL J093045.4+165930 &     0.180 &      0.177 &    0.180 &   5.84 &   9.58 &   15.81 &  17.07 &     $-$23.62 &   0.19 &   4.15e+43 &   1.82e+14 \\
SWCL J095257.1+102440 &     0.400 &      0.398 &    0.400 &   4.07 &  11.79 &   29.35 &  18.83 &     $-$23.92 &   0.02 &   1.52e+44 &   3.55e+14 \\
SWCL J095513.4+181215 &     0.423 &      0.421 &    0.423 &   4.57 &  11.61 &   33.60 &  19.54 &     $-$23.33 &   0.57 &   1.34e+44 &   3.23e+14 \\
SWCL J095515.5+180357 &     0.745 &      0.747 &    0.745 &   8.29 &  11.16 &  962.85 &  20.37 &     $-$25.03 &   0.19 &   3.47e+44 &   4.68e+14 \\
SWCL J101341.5+430651 &     0.400 &      0.400 &    0.400 &   4.09 &  11.11 &   26.72 &  18.30 &     $-$24.53 &   0.03 &   1.37e+44 &   3.34e+14 \\
SWCL J111736.0+033711 &     0.403 &      0.400 &    0.403 &   4.40 &  12.78 &   35.37 &  19.67 &     $-$23.46 &   0.26 &   3.71e+44 &   6.17e+14 \\
SWCL J113427.6$-$070208 &     0.239 &      0.304 &    0.239 &   5.02 &  10.36 &   17.31 &  17.44 &     $-$24.04 &   0.01 &   1.21e+44 &   3.41e+14 \\
SWCL J114232.3+505623 &     0.254 &   \nodata  &    0.254 &   3.47 &   9.06 &   15.30 &  16.77 &     $-$24.81 &   0.02 &   8.59e+43 &   2.74e+14 \\
SWCL J114503.1+600811 &     0.268 &      0.262 &    0.268 &   5.61 &  14.15 &   23.77 &  18.31 &     $-$23.34 &   0.01 &   5.52e+43 &   2.06e+14 \\
SWCL J114553.0+595320 &  \nodata  &      0.546 &    0.546 &   4.53 &   4.85 &   45.29 &  21.27 &     $-$22.63 &   0.36 &   1.39e+44 &   3.04e+14 \\
SWCL J115811.3+452903 &     0.389 &      0.407 &    0.389 &   6.05 &  17.12 &   39.64 &  18.39 &     $-$24.48 &   0.22 &   1.04e+44 &   2.83e+14 \\
SWCL J120137.8+104936 &     0.277 &      0.745 &    0.277 &   3.37 &   7.66 &   13.36 &  17.82 &     $-$23.78 &   1.16 &   1.39e+44 &   3.63e+14 \\
SWCL J121628.2+353820 &     0.347 &      0.358 &    0.358 &   4.52 &   8.68 &   25.13 &  19.57 &     $-$23.74 &   0.81 &   1.12e+44 &   3.03e+14 \\
SWCL J121711.8+353745 &     0.492 &      0.472 &    0.492 &   3.89 &   8.83 &   37.41 &  20.33 &     $-$23.00 &   0.25 &   2.30e+44 &   4.32e+14 \\
SWCL J123313.9+210217 &     0.516 &      0.516 &    0.516 &   6.51 &  13.68 &   69.95 &  19.74 &     $-$24.07 &   0.11 &   2.26e+44 &   4.20e+14 \\
SWCL J123612.4+290222 &     0.104 &      0.105 &    0.104 &   5.21 &   7.07 &   16.07 &  16.09 &     $-$23.29 &   0.21 &   7.48e+43 &   2.75e+14 \\
SWCL J124308.5+170639 &     0.722 &      0.145 &    0.145 &   6.12 &   7.85 &   17.97 &  15.52 &     $-$25.00 &   0.30 &   3.71e+43 &   1.74e+14 \\
SWCL J124312.1+170454 &     0.132 &      0.136 &    0.136 &   6.84 &   7.23 &   16.93 &  15.52 &     $-$24.99 &   0.20 &   5.65e+43 &   2.27e+14 \\
SWCL J125957.2+155717 &     0.700 &      0.685 &    0.700 &   4.74 &   6.82 &  296.56 &  21.28 &     $-$23.71 &   0.62 &   2.83e+44 &   4.25e+14 \\
SWCL J130332.1+591556 &     0.218 &      0.147 &    0.218 &   4.81 &   9.70 &   15.75 &  17.37 &     $-$23.56 &   0.42 &   1.06e+44 &   3.20e+14 \\
SWCL J130911.8+611521 &     0.253 &   \nodata  &    0.253 &   3.67 &   7.88 &   13.44 &  12.55 &     $-$29.00 &   0.01 &   8.34e+43 &   2.69e+14 \\
SWCL J130959.1+612530 &     0.248 &      0.247 &    0.248 &   6.19 &  12.52 &   20.77 &  17.85 &     $-$23.44 &   0.04 &   1.05e+44 &   3.12e+14 \\
SWCL J131521.9+164155 &     0.342 &      0.242 &    0.242 &   4.79 &   5.97 &   13.81 &  16.35 &     $-$24.70 &   0.01 &   3.78e+44 &   6.93e+14 \\
SWCL J133051.0+420641 &     0.597 &      0.588 &    0.588 &   4.93 &   6.65 &   86.40 &  20.37 &     $-$23.82 &   0.60 &   1.70e+44 &   3.35e+14 \\
SWCL J133055.8+420015 &  \nodata  &      0.113 &    0.113 &   3.92 &   4.31 &   11.24 &  14.57 &     $-$23.36 &   0.00 &   1.13e+44 &   3.54e+14 \\
SWCL J135914.0+470528 &     0.526 &      0.527 &    0.526 &   8.86 &  17.97 &   95.73 &  20.55 &     $-$23.32 &   0.28 &   1.03e+44 &   2.56e+14 \\
SWCL J140637.3+274348 &     0.600 &      0.758 &    0.600 &   5.34 &  10.78 &  125.86 &  20.08 &     $-$24.41 &   0.17 &   1.35e+45 &   1.20e+15 \\
SWCL J140639.0+273546 &     0.232 &      0.233 &    0.232 &   5.95 &  14.44 &   24.57 &  18.37 &     $-$22.67 &   0.07 &   1.39e+44 &   3.75e+14 \\
SWCL J140726.4+274738 &     0.177 &      0.174 &    0.174 &   4.46 &  10.17 &   22.41 &  16.06 &     $-$24.26 &   0.01 &   7.85e+43 &   2.72e+14 \\
SWCL J140907.4+242406 &     0.566 &      0.579 &    0.566 &   4.66 &   9.67 &   72.16 &  20.32 &     $-$23.58 &   0.44 &   1.24e+44 &   2.80e+14 \\
SWCL J141221.8+165216 &     0.571 &      0.438 &    0.571 &   4.32 &   7.80 &   65.16 &  21.09 &     $-$23.32 &   0.43 &   2.25e+44 &   4.04e+14 \\
SWCL J143211.6+362225 &     0.663 &      0.623 &    0.663 &   4.75 &   9.94 &  225.21 &  21.34 &     $-$23.47 &   0.11 &   1.63e+44 &   3.10e+14 \\
SWCL J144209.2+333414 &     0.451 &      0.638 &    0.638 &   6.42 &   6.84 &  151.86 &  22.04 &     $-$23.64 &   0.85 &   3.31e+44 &   4.90e+14 \\
SWCL J144604.5+203334 &     0.275 &   \nodata  &    0.275 &   3.19 &   6.77 &   12.01 &  19.09 &     $-$22.68 &   0.17 &   8.84e+43 &   2.75e+14 \\
SWCL J151550.9+442056 &     0.270 &      0.109 &    0.109 &   5.44 &   7.48 &   19.24 &  16.39 &     $-$23.00 &   0.02 &   1.04e+44 &   3.36e+14 \\
SWCL J152252.9+253527 &     0.546 &      0.626 &    0.546 &   7.20 &  13.06 &   89.91 &  19.59 &     $-$24.67 &   0.01 &   5.83e+44 &   7.41e+14 \\
SWCL J152316.4+254754 &  \nodata  &      0.796 &    0.796 &   7.60 &   3.03 & 1156.89 &  21.59 &     $-$24.16 &   0.84 &   1.12e+44 &   2.24e+14 \\
SWCL J155117.4+445118 &     0.697 &   \nodata  &    0.697 &   5.66 &   9.43 &  328.74 &  21.99 &     $-$22.95 &   0.58 &   3.76e+44 &   5.09e+14 \\
SWCL J155159.8+445748 &     0.213 &      0.207 &    0.213 &   6.23 &  12.02 &   20.09 &  16.60 &     $-$24.27 &   0.01 &   7.90e+43 &   2.67e+14 \\
SWCL J155555.3+410548 &     0.368 &   \nodata  &    0.368 &   5.10 &  14.71 &   32.07 &  18.17 &     $-$24.29 &   0.01 &   3.32e+44 &   5.89e+14 \\
SWCL J155708.6+354100 &     0.429 &      0.440 &    0.429 &   4.20 &  10.80 &   30.22 &  19.66 &     $-$23.50 &   0.15 &   1.10e+44 &   2.85e+14 \\
SWCL J155743.3+353020 &     0.166 &      0.161 &    0.166 &  35.68 &  54.31 &   97.45 &  15.91 &     $-$24.56 &   0.04 &   8.77e+44 &   1.22e+15 \\
SWCL J160637.5+321351 &     0.065 &      0.058 &    0.065 &   5.22 &   7.27 &   25.61 &  16.22 &     $-$22.02 &   0.14 &   7.15e+43 &   2.73e+14 \\
SWCL J160956.9+301052 &     0.114 &      0.276 &    0.114 &   4.64 &  10.02 &   21.14 &  17.66 &     $-$21.55 &   0.17 &   4.36e+43 &   1.95e+14 \\
SWCL J164637.4+363021 &  \nodata  &      0.077 &    0.077 &   3.54 &   2.93 &    9.67 &  17.39 &     $-$21.02 &   0.17 &   2.12e+43 &   1.28e+14 \\
SWCL J164956.4+313021 &     0.734 &   \nodata  &    0.734 &   4.82 &   8.39 &  617.61 &  21.73 &     $-$23.83 &   0.59 &   4.32e+44 &   5.41e+14 \\
SWCL J170542.2+112451 &  \nodata  &      0.409 &    0.409 &   3.69 &   6.25 &   25.12 &  19.83 &     $-$23.19 &   0.59 &   3.12e+44 &   5.52e+14 \\
SWCL J170716.1+235208 &     0.084 &      0.085 &    0.085 &   4.83 &   4.79 &   15.17 &  20.35 &     $-$22.59 &   0.28 &   3.88e+43 &   1.85e+14 \\
SWCL J170757.1+235135 &     0.212 &      0.213 &    0.212 &   6.34 &  10.28 &   17.97 &  17.64 &     $-$23.43 &   0.01 &   2.47e+44 &   5.42e+14 \\
SWCL J172011.1+692315 &     0.470 &      0.270 &    0.270 &   4.90 &   6.77 &   15.61 &  20.80 &     $-$22.34 &   0.72 &   6.44e+43 &   2.27e+14 \\
SWCL J200031.1+085259 &  \nodata  &      0.047 &    0.047 &   3.94 &   3.28 &    2.21 &  18.63 &     $-$19.87 &   0.15 &   8.29e+41 &   1.73e+13 \\
SWCL J215411.4$-$001127 &     0.381 &      0.214 &    0.381 &   6.14 &  14.54 &   41.65 &  20.52 &     $-$21.98 &   0.20 &   1.69e+44 &   3.84e+14 \\
SWCL J215411.6$-$000654 &     0.213 &      0.106 &    0.106 &   3.68 &   2.77 &    7.72 &  18.60 &     $-$20.85 &   0.10 &   4.34e+43 &   1.96e+14 \\
SWCL J215413.2+000413 &     0.227 &      0.166 &    0.227 &   4.60 &  10.58 &   18.72 &  17.67 &     $-$23.65 &   0.04 &   1.70e+44 &   4.25e+14 \\
SWCL J215423.1+000526 &     0.169 &      0.170 &    0.169 &   8.58 &  15.76 &   28.35 &  16.10 &     $-$24.23 &   0.07 &   1.64e+44 &   4.31e+14 \\
SWCL J215520.0$-$001115 &     0.373 &      0.371 &    0.371 &   3.83 &   7.08 &   26.74 &  19.81 &     $-$23.70 &   0.85 &   1.16e+44 &   3.06e+14 \\
SWCL J222432.9$-$021216 &     0.204 &      0.162 &    0.162 &   4.00 &   5.54 &   13.21 &  18.57 &     $-$21.52 &   0.25 &   4.69e+43 &   1.99e+14 \\
SWCL J222438.0$-$022231 &     0.499 &      0.496 &    0.499 &   6.44 &  13.73 &   71.02 &  19.91 &     $-$23.83 &   0.06 &   1.83e+44 &   3.72e+14 \\
SWCL J222439.0$-$021111 &  \nodata  &      0.292 &    0.292 &   5.16 &   8.99 &   24.41 &  18.44 &     $-$23.89 &   0.56 &   4.75e+43 &   1.85e+14 \\
SWCL J222444.0$-$022034 &     0.714 &      0.718 &    0.714 &   3.84 &   6.71 &  546.21 &  22.08 &     $-$23.29 &   0.44 &   2.81e+44 &   4.20e+14 \\
SWCL J222506.2$-$020611 &     0.524 &   \nodata  &    0.524 &   3.45 &   6.35 &   40.70 &  20.50 &     $-$23.20 &   0.75 &   1.47e+44 &   3.20e+14 \\
SWCL J222516.4$-$020825 &     0.467 &      0.585 &    0.467 &   3.56 &   8.61 &   34.29 &  20.42 &     $-$22.99 &   0.60 &   1.74e+44 &   3.69e+14 \\
SWCL J222954.1+194350 &     0.287 &      0.290 &    0.287 &   7.65 &  15.63 &   28.27 &  18.00 &     $-$23.63 &   0.03 &   1.12e+44 &   3.15e+14 \\
SWCL J224206.9+233408 &     0.402 &   \nodata  &    0.402 &   5.38 &  13.98 &   40.30 &  19.40 &     $-$23.76 &   0.13 &   3.19e+44 &   5.61e+14 \\
SWCL J231257.7+182543 &     0.325 &      0.548 &    0.325 &   3.83 &   8.73 &   21.26 &  19.98 &     $-$22.50 &   0.54 &   1.07e+44 &   2.99e+14 \\
SWCL J231733.7+322828 &     0.400 &      0.376 &    0.400 &   4.35 &  13.24 &   37.54 &  19.32 &     $-$23.25 &   0.07 &   1.61e+44 &   3.67e+14 \\
SWCL J232244.2+055601 &     0.235 &   \nodata  &    0.235 &   4.18 &   7.80 &   13.89 &  17.51 &     $-$24.04 &   0.31 &   5.55e+43 &   2.11e+14 \\
SWCL J232248.4+054810 &     0.320 &      0.244 &    0.244 &   5.92 &   9.94 &   23.97 &  16.85 &     $-$24.56 &   0.01 &   1.72e+44 &   4.24e+14 \\
SWCL J232717.2+263108 &     0.223 &      0.223 &    0.223 &  11.50 &  13.29 &   31.74 &  16.85 &     $-$24.33 &   0.05 &   1.41e+44 &   3.80e+14 \\
SWCL J232725.6+263506 &     0.257 &      0.227 &    0.257 &   5.71 &   8.79 &   16.05 &  17.99 &     $-$23.13 &   0.38 &   1.72e+44 &   4.21e+14 \\
SWCL J233009.3+264459 &  \nodata  &      0.522 &    0.522 &   3.68 &   4.00 &   36.17 &  19.82 &     $-$23.99 &   0.56 &   1.07e+44 &   2.64e+14 \\

\enddata
\tablecomments{If the source region size (2\arcmin or 3\arcmin) confirmed a redshift, the value is given. If not, then the non-detection is labelled with '...'. Cases where neither source region size confirmed a redshift are not included in this table.  Confirmed SDSS galaxy over-densities are listed as well as the significance of the detection. Cases where the 2\arcmin\ and 3\arcmin\ regions' redshifts differ are discussed in Section \ref{sec:method}. $N_{Net}$ is the number of galaxies above the background count for the detected redshift bin. $N_{opt}$ is the estimated optical richness, described in Section number \ref{subsec:opticalprop}. We also list the BCG SDSS$r$ apparent magnitude, the calculated absolute magnitude  ($r_{abs}$) and the BCG-to-X-ray offset in Mpc. $L_X$ is the X-ray bolometric luminosity and $M_{500}$ is a mass estimate similar to the virialized mass. Units are given in column headings in parenthesis.}
\end{deluxetable}
\end{landscape}
\pagestyle{plain}

\begin{deluxetable}{cccccccccccc}

\tabletypesize{\scriptsize}
\tablecolumns{12}
\tablewidth{0pt}
\tablecaption{2\arcmin\ and 3\arcmin\ Regions Confirmed, But Different Redshifts \label{tab:cases}}

\tablehead{
\colhead{\swift\ Name} & \colhead{$z_2$} & \colhead{$z_3$} & \colhead{$\sigma_2$} & \colhead{$\sigma_3$} & \colhead{Offset$_2$} & \colhead{Offset$_3$} & \colhead{Score} & 
}
\startdata

SWCL J002114.5+205943 & \textbf{      0.141} &       0.074 &       6.075 &       5.425 &     0.056 &     0.108 &     0.269 & \\
SWCL J035259.4$-$004338 &       0.413 & \textbf{      0.301} &       4.398 &       3.775 &     0.358 &     0.004 &    $-$0.145 & \\
SWCL J054716.7+641156 &       0.271 & \textbf{      0.369} &       3.193 &       6.190 &     0.282 &     0.797 &    $-$0.485 & \\
SWCL J075900.8+324449 & \textbf{      0.579} &       0.751 &       3.696 &       3.017 &     0.644 &     0.838 &     0.421 & \\
SWCL J092719.6+301348 &       0.396 & \textbf{      0.331} &       3.717 &       4.255 &     0.522 &     0.037 &    $-$0.664 & \\
SWCL J092852.3+002137 &       0.535 & \textbf{      0.091} &       3.443 &       4.092 &     0.604 &     0.022 &    $-$0.799 & \\
SWCL J113427.6$-$070208 & \textbf{      0.239} &       0.304 &       5.024 &       4.663 &     0.010 &     0.013 &     0.122 & \\
SWCL J120137.8+104936 & \textbf{      0.277} &       0.745 &       3.367 &       4.565 &     0.085 &     1.159 &     0.675 & \\
SWCL J124308.5+170639 &       0.722 & \textbf{      0.145} &       3.898 &       6.120 &     0.601 &     0.303 &    $-$1.038 & \\
SWCL J130332.1+591556 & \textbf{      0.218} &       0.147 &       4.808 &       4.112 &     0.422 &     0.332 &     0.142 & \\
SWCL J131521.9+164155 &       0.342 & \textbf{      0.242} &       5.038 &       4.786 &     0.533 &     0.014 &    $-$0.435 & \\
SWCL J140637.3+274348 & \textbf{      0.600} &       0.758 &       5.342 &       4.929 &     0.168 &     0.545 &     0.515 & \\
SWCL J141221.8+165216 & \textbf{      0.571} &       0.438 &       4.325 &       3.997 &     0.434 &     0.860 &     0.535 & \\
SWCL J144209.2+333414 &       0.451 & \textbf{      0.638} &       3.256 &       6.423 &     0.314 &     0.851 &    $-$0.519 & \\
SWCL J151550.9+442056 &       0.270 & \textbf{      0.109} &       3.903 &       5.444 &     0.152 &     0.017 &    $-$0.648 & \\
SWCL J152252.9+253527 & \textbf{      0.546} &       0.626 &       7.201 &       5.655 &     0.014 &     0.015 &     0.516 & \\
SWCL J160956.9+301052 & \textbf{      0.114} &       0.276 &       4.639 &       3.975 &     0.173 &     0.362 &     0.410 & \\
SWCL J172011.1+692315 &       0.470 & \textbf{      0.270} &       3.066 &       4.902 &     0.695 &     0.724 &    $-$0.583 & \\
SWCL J215411.4$-$001127 & \textbf{      0.381} &       0.214 &       6.142 &       4.248 &     0.202 &     0.446 &     0.876 & \\
SWCL J215411.6$-$000654 &       0.213 & \textbf{      0.106} &       3.542 &       3.680 &     0.211 &     0.098 &    $-$0.159 & \\
SWCL J215413.2+000413 &       0.227 & \textbf{      0.166} &       4.597 &       8.218 &     0.041 &     0.460 &    $-$0.788 & \\
SWCL J222516.4$-$020825 & \textbf{      0.467} &       0.585 &       3.564 &       3.523 &     0.602 &     0.996 &     0.407 & \\
SWCL J231257.7+182543 & \textbf{      0.325} &       0.548 &       3.832 &       3.597 &     0.543 &     0.530 &     0.065 & \\
SWCL J232248.4+054810 &       0.320 & \textbf{      0.244} &       4.636 &       5.925 &     0.080 &     0.009 &    $-$0.501 & \\

\enddata
\tablecomments{$z_2$, $z_3$: redshifts confirmed by the 2\arcmin\ and 3\arcmin\ regions respectively, bold represents the final redshift determined for that candidate.  $\sigma_2$, $\sigma_3$: standard deviations above the average.  Offset$_2$, Offset$_3$: offsets in Mpc between BCG and X-ray source's center.  Score =  $(\sigma_2 - \sigma_3)/3 - (\text{Offset}_2 - \text{Offset}_3)/$Mpc.  If Score $>$ ($<$) 0 then 2\arcmin\ (3\arcmin) region confirmed. }

\end{deluxetable}

\pagestyle{empty}
\begin{landscape}
\begin{deluxetable}{ccccccccccc}

\tabletypesize{\scriptsize}
\tablecolumns{11}
\tablewidth{0pt}
\tablecaption{Close Contamination Cases \label{tab:close}}
\tablehead{
\colhead{Name 1} & \colhead{Name 2} & \colhead{Distance} & \colhead{$z_1$} & \colhead{$z_2$} & \colhead{$\sigma_1$} & \colhead{$\sigma_2$} & \colhead{BCG$_1$} & \colhead{BCG$_2$} & \colhead{Multiple peak?} & \colhead{Conclusion}  
}
\startdata

 SWCL J085552.1 & SWCL J085619.7 & 4.799 & 0.477 & 0.479 & 4.01 & 3.53 & 0.694 & 0.050 &  no & 1: cluster at 0.477, 2: $z$ = 0.479            \\
 SWCL J092730.1 & SWCL J092719.6 & 3.795 & 0.303 & 0.331 & 4.41 & 4.25 & 0.254 & 0.036 &  no & 1: cluster at $z$ = 0.303, 2: $z$ = 0.331        \\
 SWCL J215423.1 & SWCL J215413.2 & 2.745 & 0.169 & 0.166 & 8.58 & 8.22 & 0.063 & 0.449 & yes & 1: strong cluster at 0.169, 2: $z$ = 0.227     \\
 SWCL J232725.6 & SWCL J232717.2 & 4.378 & 0.227 & 0.223 & 6.42 &11.50 & 0.328 & 0.048 & yes & 1: cluster at $z$ = 0.257, 2: strong cluster at $z$ = 0.223 \\

\enddata
\tablecomments{Close contaminations are defined to be the cases where the X-ray source centers are within 6\arcmin\ and the difference in redshifts is $\delta$$z$$\lesssim 0.05$. Distances are given in arcminutes and the BCG offsets are given in Mpc. We discuss this more in Section \ref{subsec:close}. }
\end{deluxetable}
\end{landscape}
\pagestyle{plain}

\begin{deluxetable}{cc}

\tabletypesize{\scriptsize}
\tablecolumns{10}
\tablewidth{0pt}
\tablecaption{Red Sequence Colors \label{tab:colredrange}}
\tablehead{
\colhead{Redshift Range} & \colhead{Color} }
\startdata

0.00 - 0.43 & g $-$ r \\
0.43 - 0.70 & r $-$ i \\
0.70 - 1.00 & i $-$ z \\

\enddata
\tablecomments{From Table 2 of \citet{gmbcg}.  These are the colors used for the red sequence plots, dependent on redshift.}
\end{deluxetable}

\begin{deluxetable}{cccccccccc}

\tabletypesize{\scriptsize}
\tablecolumns{10}
\tablewidth{0pt}
\tablecaption{Optical Catalogs vs Our Catalog \label{tab:gmbcg}}

\tablehead{
\colhead{\swift} & \colhead{Separation}  & \colhead{Wen} & \colhead{Wen} & \colhead{Wen} & \colhead{GMBCG} & \colhead{\swift} & 
\\
\colhead{Name     } & \colhead{(Arcmin)} & \colhead{RA} & \colhead{DEC} & \colhead{phot. z} & \colhead{phot. z} & \colhead{phot. z} &
}
\startdata

SWCL J015752.9+165933  &        0.036  &       29.471  &       16.992  &        0.507  &         \nodata    &        0.510 \\ 
SWCL J015803.5+165005  &        0.036  &       29.515  &       16.835  &        0.209  &         \nodata    &        0.223 \\ 
SWCL J020003.8+084024  &        0.101  &       30.014  &        8.674  &        0.215  &         \nodata    &        0.196 \\ 
SWCL J025630.7+000601  &        0.049  &       44.129  &        0.101  &        0.355  &        0.375  &        0.374 \\ 
SWCL J035259.4$-$004338  &        0.017  &       58.247  &       $-$0.727  &        0.325  &        0.314  &        0.301 \\ 
SWCL J054716.7+641156  &        0.932  &       86.793  &       64.209  &        0.363  &         \nodata    &        0.369 \\ 
SWCL J083340.9+331118  &        0.513  &      128.421  &       33.197  &        0.709  &         \nodata    &        0.682 \\ 
SWCL J084749.4+133142  &        0.034  &      131.956  &       13.528  &        0.359  &         \nodata    &        0.391 \\ 
SWCL J090714.8+351020  &        0.566  &      136.802  &       35.167  &        0.210  &         \nodata    &        0.218 \\ 
SWCL J092730.1+301046  &        0.789  &      141.864  &       30.188  &        0.373  &        0.293  &        0.303 \\ 
SWCL J095257.1+102440  &        0.061  &      148.237  &       10.410  &        0.375  &         \nodata    &        0.400 \\ 
SWCL J095513.4+181215  &        0.324  &      148.808  &       18.209  &        0.327  &        0.316  &        0.423 \\ 
SWCL J095515.5+180357  &        0.983  &      148.818  &       18.050  &        0.622  &         \nodata    &        0.745 \\ 
SWCL J101341.5+430651  &        0.097  &      153.424  &       43.116  &        0.423  &         \nodata    &        0.400 \\ 
SWCL J111736.0+033711  &        0.817  &      169.387  &        3.625  &        0.442  &         \nodata    &        0.403 \\ 
SWCL J113427.6$-$070208  &        0.046  &      173.616  &       $-$7.036  &        0.251  &         \nodata    &        0.239 \\ 
SWCL J114232.3+505623  &        0.069  &      175.636  &       50.940  &        0.260  &         \nodata    &        0.254 \\ 
SWCL J114553.0+595320  &        0.010  &      176.471  &       59.889  &        0.182  &         \nodata    &        0.546 \\ 
SWCL J115811.3+452903  &        0.701  &      179.560  &       45.492  &        0.409  &        0.416  &        0.389 \\ 
SWCL J121628.2+353820  &        0.058* &      184.117* &       35.640* &         \nodata    &        0.266  &        0.358 \\
SWCL J121711.8+353745  &        0.458  &      184.306  &       35.635  &        0.443  &        0.459  &        0.492 \\ 
SWCL J123313.9+210217  &        0.288  &      188.312  &       21.040  &        0.544  &         \nodata    &        0.516 \\ 
SWCL J130959.1+612530  &        0.166  &      197.492  &       61.423  &        0.243  &        0.254  &        0.248 \\ 
SWCL J131521.9+164155  &        0.062  &      198.840  &       16.699  &        0.223  &         \nodata    &        0.242 \\ 
SWCL J140637.3+274348  &        0.054  &      211.655  &       27.731  &        0.655  &         \nodata    &        0.600 \\ 
SWCL J140639.0+273546  &        0.232  &      211.664  &       27.600  &        0.263  &        0.259  &        0.232 \\ 
SWCL J140726.4+274738  &        0.083  &      211.861  &       27.795  &        0.159  &        0.171  &        0.174 \\ 
SWCL J143211.6+362225  &        0.845  &      218.041  &       36.361  &        0.572  &         \nodata    &        0.663 \\ 
SWCL J151550.9+442056  &        0.145  &      228.963  &       44.347  &        0.116  &         \nodata    &        0.109 \\ 
SWCL J152252.9+253527  &        0.037  &      230.720  &       25.591  &        0.549  &         \nodata    &        0.546 \\ 
SWCL J155159.8+445748  &        0.050  &      238.000  &       44.963  &        0.201  &        0.208  &        0.213 \\ 
SWCL J155555.3+410548  &        0.046  &      238.980  &       41.097  &        0.355  &        0.375  &        0.368 \\ 
SWCL J155708.6+354100  &        0.054  &      239.285  &       35.683  &        0.413  &        0.390  &        0.429 \\ 
SWCL J155743.3+353020  &        0.245  &      239.427  &       35.508  &        0.148  &        0.169  &        0.166 \\ 
SWCL J170757.1+235135  &        0.036  &      256.989  &       23.859  &        0.221  &         \nodata    &        0.212 \\ 
SWCL J215423.1+000526  &        0.381  &      328.598  &        0.084  &        0.175  &        0.156  &        0.169 \\ 
SWCL J222432.9$-$021216  &        0.378  &      336.131  &       $-$2.203  &        0.242  &         \nodata    &        0.162 \\ 
SWCL J222438.0$-$022231  &        0.164  &      336.158  &       $-$2.372  &        0.507  &         \nodata    &        0.499 \\ 
SWCL J222444.0$-$022034  &        0.061  &      336.184  &       $-$2.343  &        0.658  &         \nodata    &        0.714 \\ 
SWCL J222954.1+194350  &        0.132  &      337.473  &       19.731  &        0.272  &         \nodata    &        0.287 \\ 
SWCL J224206.9+233408  &        0.391  &      340.522  &       23.571  &        0.432  &         \nodata    &        0.402 \\ 
SWCL J231257.7+182543  &        0.983  &      348.233  &       18.414  &        0.460  &         \nodata    &        0.325 \\ 
SWCL J231733.7+322828  &        0.216  &      349.394  &       32.476  &        0.405  &         \nodata    &        0.400 \\ 
SWCL J232248.4+054810  &        0.039  &      350.701  &        5.802  &        0.244  &         \nodata    &        0.244 \\ 
SWCL J232717.2+263108  &        0.225  &      351.825  &       26.522  &        0.230  &         \nodata    &        0.223 \\ 

\enddata
\tablecomments{\ List of 45 position matches from comparing clusters of the GMBCG catalog \citep{gmbcg} and catalog of \citet{wen} to the SDSS identifications of the \swift\ AGN and Cluster Survey. We used a matching radius of 1\arcmin. Of these 45, 42 agree in redshift to $\delta z < 0.1$. The remaining three are different enough in redshift to be different clusters and thus, not a match. Note: SWCL J121628.2+353820 is the only match in the GMBCG catalog that does not have a corresponding entry in the \citet{wen} catalog, thus the separation, RA and Dec reported are actually from the GMBCG catalog. }
\end{deluxetable}

\begin{deluxetable}{cccccccccc}

\tabletypesize{\scriptsize}
\tablecolumns{10}
\tablewidth{0pt}
\tablecaption{SWXCS Catalog vs Our Catalog \label{tab:liu}}

\tablehead{
\colhead{\swift} & \colhead{Separation}  & \colhead{SWXCS} & \colhead{SWXCS} & \colhead{SWXCS} & \colhead{\swift} & 
\\
\colhead{Name     } & \colhead{(Arcmin)} & \colhead{RA} & \colhead{DEC} & \colhead{phot. z} & \colhead{phot. z} &
}
\startdata

SWCL J002114.5+205943  &        0.051  &       5.3095  &      20.9956  &         \nodata    &        0.141 \\ 
SWCL J002823.6+092705  &        0.045  &       7.0987  &       9.4517  &        0.224  &        0.195 \\ 
SWCL J012303.8+375609  &        0.150  &      20.7625  &      37.9361  &         \nodata    &         \nodata   \\ 
SWCL J015752.9+165933  &        0.018  &      29.4699  &      16.9924  &        0.507  &        0.510 \\ 
SWCL J020003.8+084024  &        0.026  &      30.0158  &       8.6730  &        0.215  &        0.196 \\ 
SWCL J020745.0+002053  &        0.098  &      31.9375  &       0.3498  &         \nodata    &         \nodata   \\ 
SWCL J035259.4$-$004338  &        0.033  &      58.2471  &      $-$0.7274  &        0.328  &        0.301 \\ 
SWCL J075900.8+324449  &        0.065  &     119.7527  &      32.7480  &         \nodata    &        0.579 \\ 
SWCL J084749.4+133142  &        0.096  &     131.9544  &      13.5276  &        0.349  &        0.391 \\ 
SWCL J092649.8+301346  &        0.043  &     141.7085  &      30.2296  &        0.559  &        0.567 \\ 
SWCL J092719.6+301348  &        0.086  &     141.8327  &      30.2309  &        0.302  &        0.331 \\ 
SWCL J092730.1+301046  &        0.058  &     141.8744  &      30.1798  &        0.312  &        0.303 \\ 
SWCL J095206.7+102137  &        0.108  &     148.0280  &      10.3622  &         \nodata    &         \nodata   \\ 
SWCL J095513.4+181215  &        0.052  &     148.8058  &      18.2051  &        0.416  &        0.423 \\ 
SWCL J101341.5+430651  &        0.058  &     153.4235  &      43.1152  &        0.449  &        0.400 \\ 
SWCL J114232.3+505623  &        0.037  &     175.6344  &      50.9393  &        0.260  &        0.254 \\ 
SWCL J114503.1+600811  &        0.048  &     176.2613  &      60.1362  &        0.285  &        0.268 \\ 
SWCL J124312.1+170454  &        0.126  &     190.8026  &      17.0812  &        0.142  &        0.136 \\ 
SWCL J133051.0+420641  &        0.150  &     202.7158  &      42.1111  &         \nodata    &        0.588 \\ 
SWCL J133055.8+420015  &        0.076  &     202.7310  &      42.0048  &         \nodata    &        0.113 \\ 
SWCL J140637.3+274348  &        0.049  &     211.6547  &      27.7302  &        0.655  &        0.600 \\ 
SWCL J140639.0+273546  &        0.109  &     211.6611  &      27.5974  &        0.252  &        0.232 \\ 
SWCL J143211.6+362225  &        0.084  &     218.0467  &      36.3740  &        0.572  &        0.663 \\ 
SWCL J143223.3+361752  &        0.028  &     218.0967  &      36.2980  &         \nodata    &         \nodata   \\ 
SWCL J152252.9+253527  &        0.152  &     230.7187  &      25.5926  &        0.557  &        0.546 \\ 
SWCL J155117.4+445118  &        0.083  &     237.8208  &      44.8551  &         \nodata    &        0.697 \\ 
SWCL J155743.3+353020  &        0.082  &     239.4292  &      35.5065  &        0.158  &        0.166 \\ 
SWCL J164956.4+313021  &        0.082  &     252.4857  &      31.5070  &         \nodata    &        0.734 \\ 
SWCL J173932.8+272051  &        0.038  &     264.8860  &      27.3480  &         \nodata    &         \nodata   \\ 
SWCL J194004.2+782415  &        0.051  &     295.0164  &      78.4033  &         \nodata    &         \nodata   \\ 
SWCL J215507.7+164725  &        0.028  &     328.7821  &      16.7907  &         \nodata    &         \nodata   \\ 
SWCL J222438.0$-$022231  &        0.092  &     336.1594  &      $-$2.3742  &        0.507  &        0.499 \\ 
SWCL J222444.0$-$022034  &        0.006  &     336.1832  &      $-$2.3428  &        0.658  &        0.714 \\ 
SWCL J222516.4$-$020825  &        0.068  &     336.3192  &      $-$2.1411  &         \nodata    &        0.467 \\ 
SWCL J222954.1+194350  &        0.159  &     337.4729  &      19.7292  &        0.272  &        0.287 \\ 
SWCL J232248.4+054810  &        0.053  &     350.7014  &       5.8036  &        0.244  &        0.244 \\ 
SWCL J232725.6+263506  &        0.037  &     351.8565  &      26.5843  &        0.059  &        0.257 \\ 

\enddata
\tablecomments{\ List of 37 position matches comparing clusters of the SWXCS catalog \citep{liu} to the SDSS identifications of the \swift\ AGN and Cluster Survey. Both cluster surveys use \swift\ fields to locate extended X-ray sources as potential galaxy clusters. We used a matching radius of 10\arcsec. The redshifts of SWXCS are reported from various optical, X-ray and SZ catalogs (see \citet{liu} for details). For SWXCS clusters with multiple redshifts listed, we report the average of these redshifts here. Of these 37 position matches, 23 have redshifts in both catalogs. 22 of 23 clusters agree in redshift to $\delta z < 0.1$. The remaining position match is different enough in redshift to be considered different clusters and thus, not a match. 19 clusters agree within redshift $\delta z < 0.05$.}
\end{deluxetable}

\pagestyle{empty}
\begin{landscape}
\begin{deluxetable}{rcccccccccccc}
\tabletypesize{\scriptsize}
\tablecolumns{10}
\tablewidth{0pt}
\tablecaption{SDSS Confirmations of Lower Significance SACS Clusters: \label{tab:notcatalog}}
\tablehead{
 \colhead{\swift} & \colhead{2\arcmin\ Reg.} & \colhead{3\arcmin\ Reg.} & \colhead{Conf.} & \colhead{Conf.}
& \colhead{$N_{Net}$} & \colhead{$N_{opt}$} & \colhead{BCG} & \colhead{BCG} & \colhead{Offset} & \colhead{$L_X$} & \colhead{$M_{500}$}
\\
 \colhead{Name} & \colhead{Conf. $z$} & \colhead{Conf. $z$} & \colhead{Phot. $z$} & \colhead{$\sigma$ } & \colhead{ } & \colhead{ } & \colhead{$r_{mag}$} & \colhead{$r_{abs}$} & \colhead{(Mpc)} & \colhead{(erg/s)} & \colhead{(M$_{\odot})$} 

}
\startdata

SWCL J015803.5+165006 & 0.223 & 0.232 &   0.223 &  7.42 & 11.92 & 3.43  & 16.95 &   $-$24.33 &  0.01 &   8.08e+43 &   2.69e+14 \\
SWCL J095513.4+181216 & 0.423 & 0.421 &   0.423 &  4.51 & 11.59 & 5.58  & 19.54 &   $-$23.34 &  0.57 &   1.29e+44 &   3.15e+14 \\
SWCL J095515.7+180359 & 0.745 & 0.747 &   0.745 &  8.35 & 11.14 & 160   & 20.37 &   $-$25.03 &  0.21 &   3.34e+44 &   4.57e+14 \\
SWCL J083340.9+331117 & 0.682 & 0.676 &   0.682 &  8.42 & 18.06 & 112   & 20.66 &   $-$24.36 &  0.22 &   5.38e+44 &   6.43e+14 \\
SWCL J124313.2+170443 & 0.132 & 0.134 &   0.132 &  6.28 & 10.16 & 3.31  & 15.80 &   $-$24.02 &  0.03 &   6.49e+43 &   2.48e+14 \\
SWCL J224206.2+233350 & 0.403 &  \nodata   &   0.403 &  5.69 & 15.04 & 7.18  & 19.40 &   $-$23.76 &  0.15 &   3.56e+44 &   6.01e+14 \\
SWCL J115811.3+452904 & 0.389 & 0.407 &   0.389 &  5.90 & 17.08 & 6.58  & 18.39 &   $-$24.48 &  0.22 &   1.08e+44 &   2.91e+14 \\
SWCL J160637.7+321349 & 0.065 & 0.064 &   0.065 &  5.22 &  7.25 & 4.23  & 16.22 &   $-$22.02 &  0.14 &   7.45e+43 &   2.80e+14 \\
SWCL J082114.0+320018 & 0.746 & 0.732 &   0.746 &  7.39 &  9.08 & 142   & 21.16 &   $-$24.28 &  0.08 &   9.15e+44 &   8.55e+14 \\
SWCL J232248.3+054810 & 0.319 & 0.245 &   0.245 &  5.63 &  9.50 & 22.8  & 16.85 &   $-$24.56 &  0.01 &   1.78e+44 &   4.33e+14 \\
SWCL J222954.1+194349 & 0.286 & 0.291 &   0.286 &  8.18 & 16.66 & 5.03  & 18.00 &   $-$23.63 &  0.04 &   1.13e+44 &   3.19e+14 \\
SWCL J143211.7+362226 & 0.663 & 0.623 &   0.663 &  4.66 & 10.36 & 39.0  & 21.34 &   $-$23.47 &  0.10 &   1.79e+44 &   3.29e+14 \\

\enddata
\tablecomments{\ Clusters we found in a previous version of our X-ray selection method, but excluded from the current catalog due to changes in the significance threshold. Columns are the same as in Table \ref{tab:confz}.}

\end{deluxetable}
\end{landscape}
\pagestyle{plain}

\clearpage

\begin{thebibliography}{}

\bibitem[Abell~\etal(1989)]{opt1} Abell, G.~O., Corwin, H.~G.,~Jr., \& Olowin, R.~P.\ 1989, \apj, 70, 1A
\bibitem[Adami et al.(2011)]{xmmlss} Adami, C., Mazure, A., Pierre, M., et al.\ 2011, A\&A, 526, A18 
\bibitem[Adelman-McCarthy et al.(2007)]{dr5} Adelman-McCarthy, J.~K., Ag{\"u}eros, M.~A., Allam, S.~S., et al.\ 2007, ApJS, 172, 634 
\bibitem[Aihara et al.(2011)]{dr8} Aihara, H., Allende Prieto, C., An, D., et al.\ 2011, ApJS, 193, 29 
\bibitem[Allen~\etal(2011)]{allen} Allen, S.~W., Evrard, A.~E., \& Mantz, A.~B.\ 2011, ARA\&A, 49, 409
\bibitem[Anderson et al.(2015)]{2015MNRAS.449.3806A} Anderson, M.~E., Gaspari, M., White, S.~D.~M., Wang, W., \& Dai, X.\ 2015, MNRAS, 449, 3806 
\bibitem[Arnaud(1996)]{xspec} Arnaud, K.~A.\ 1996, Astronomical Data Analysis Software and Systems V, 101, 17
\bibitem[Ascaso et al.(2014)]{dls} Ascaso, B., Wittman, D., \& Dawson, W.\ 2014, MNRAS, 439, 1980 
\bibitem[Assef et al.(2010)]{assef10} Assef, R.~J., Kochanek, C.~S., Brodwin, M., et al.\ 2010, \apj, 713, 970 
\bibitem[Bahcall~(1988)]{bahcall} Bahcall, N.~A.\ 1988, ARA\&A, 26, 631B
\bibitem[Barkhouse et al.(2006)]{barkhouse} Barkhouse, W.~A., Green, P.~J., Vikhlinin, A., et al.\ 2006, \apj, 645, 955 
\bibitem [Bell~\etal(2003)]{bell03} Bell, E.~F., McIntosh, D.~H., Katz, N., \& Weinberg, M.~D. \ 2003, ApJS, 149, 289B
\bibitem[Bell et al.(2004)]{bell04} Bell, E.~F., Wolf, C., Meisenheimer, K., et al.\ 2004, \apj, 608, 752
\bibitem[Blackburne \& Kochanek(2012)]{blackburne} Blackburne, J.~A., \& Kochanek, C.~S.\ 2012, \apj, 744, 76 
\bibitem[B{\"o}hringer et al.(2000)]{noras} B{\"o}hringer, H., Voges, W., Huchra, J.~P., et al.\ 2000, ApJS, 129, 435 
\bibitem[Burenin et al.(2007)]{400sd} Burenin, R.~A., Vikhlinin, A., Hornstrup, A., et al.\ 2007, ApJS, 172, 561 
\bibitem[Carlstrom et al.(2002)]{sz1} Carlstrom, J.~E., Holder, G.~P., \& Reese, E.~D.\ 2002, ARA\&A, 40, 643
\bibitem[D'Elia et al.(2013)]{delia} D'Elia, V., Perri, M., Puccetti, S., et al.\ 2013, A\&A, 551, A142 
\bibitem[Dai~\etal(2007)]{dai07} Dai, X., Kochanek, C.~S., \& Morgan, N.~D.\ 2007, \apj, 658, 917 
\bibitem[Dai et al.(2009)]{dai09} Dai, X., Assef, R.~J., Kochanek, C.~S., et al.\ 2009, \apj, 697, 506
\bibitem[Dai et al.(2010)]{dai10} Dai, X., Bregman, J.~N., Kochanek, C.~S., \& Rasia, E.\ 2010, \apj, 719, 119 
\bibitem[Dai~\etal(2015)]{swift1} Dai, X., Griffin, R.~D., Kochanek, C.~S., Nugent, J.~M., \& Bregman, J.~N.\ 2015, ApJS, 218, 8 
\bibitem[Donahue et al.(2001)]{donahue} Donahue, M., Mack, J., Scharf, C., et al.\ 2001, ApJL, 552, L93 
\bibitem[Evans et al.(2014)]{evans} Evans, P.~A., Osborne, J.~P., Beardmore, A.~P., et al.\ 2014, ApJS, 210, 8
\bibitem[Finoguenov et al.(2007)]{finoguenov} Finoguenov, A., Guzzo, L., Hasinger, G., et al.\ 2007, ApJS, 172, 182
\bibitem[Finoguenov et al.(2015)]{finoguenov15} Finoguenov, A., Tanaka, M., Cooper, M., et al.\ 2015, A\&A, 576, A130
\bibitem[Gal et al.(2003)]{opt2} Gal, R.~R., de Carvalho, R.~R., Lopes, P.~A.~A., et al.\ 2003, AJ, 125, 2064
\bibitem[Girardi et al.(1998)]{girardi} Girardi, M., Giuricin, G., Mardirossian, F., Mezzetti, M., \& Boschin, W.\ 1998, \apj, 505, 74 
\bibitem[Gladders~\&~Yee(2005)]{red1} Gladders, M.~D. \& Yee, H.~K.~C. \ 2005, ApJS, 157, 1
\bibitem[Guzzo et al.(2009)]{reflex} Guzzo, L., Schuecker, P., B{\"o}hringer, H., et al.\ 2009, A\&A, 499, 357 
\bibitem[Hao et al.(2010)]{gmbcg} Hao, J., McKay, T.~A., Koester, B.~P., et al.\ 2010, ApJS, 191, 254
\bibitem[Hasselfield et al.(2013)]{sz5} Hasselfield, M., Hilton, M., Marriage, T.~A., et al.\ 2013, JCAP, 7, 8 
\bibitem[Hoekstra (2007)]{gravlens3} Hoekstra, H.\ 2007, MNRAS, 379, 317H
\bibitem[Hoekstra \& Jain(2008)]{gravlens1} Hoekstra, H., \& Jain, B.\ 2008, Annual Review of Nuclear and Particle Science, 58, 99
\bibitem[Isobe et al.(1990)]{isobe} Isobe, T., Feigelson, E.~D., Akritas, M.~G., \& Babu, G.~J.\ 1990, \apj, 364, 104 
\bibitem[Kaiser(1991)]{1991ApJ...383..104K} Kaiser, N.\ 1991, \apj, 383, 104 
\bibitem[Kochanek et al.(2003)]{kochanek} Kochanek, C.~S., White, M., Huchra, J., et al.\ 2003, \apj, 585, 161 
\bibitem[Koester et al.(2007)]{maxbcg} Koester, B.~P., McKay, T.~A., Annis, J., et al.\ 2007, \apj, 660, 239
\bibitem[Kravtsov~\&~Borgani(2012)]{formofgal} Kravtsov, A.~V. \& Borgani, S.\ 2012, ARA\&A, 50, 353K
\bibitem[Lin~\&~Mohr(2004)]{bcg2} Lin, Y. \& Mohr, J.~J.\ 2004, \apj, 617, 879
\bibitem[Liu et al.(2015)]{liu} Liu, T., Tozzi, P., Tundo, E., et al.\ 2015, ApJS, 216, 28 
\bibitem[Lopes~\etal(2006)]{baxnosocs} Lopes, P.~A.~A., de Carvalho, R.~R., \& Capelato, H.~V. \ 2006, \apj, 648, 209L
\bibitem[Marriage et al.(2011)]{sz4} Marriage, T.~A., Acquaviva, V., Ade, P.~A.~R., et al.\ 2011, \apj, 737, 61 
\bibitem[Mehrtens et al.(2012)]{mehrtens} Mehrtens, N., Romer, A.~K., Hilton, M., et al.\ 2012, MNRAS, 423, 1024 
\bibitem[Navarro et al.(1995)]{nfw} Navarro, J.~F., Frenk, C.~S., \& White, S.~D.~M.\ 1995, MNRAS, 275, 720 
\bibitem[Nilo Castell{\'o}n et al.(2014)]{nilo} Nilo Castell{\'o}n, J.~L., Alonso, M.~V., Garc{\'{\i}}a Lambas, D., et al.\ 2014, MNRAS, 437, 2607
\bibitem[Oguri et al.(2010)]{gravlens6} Oguri, M., Takada, M., Okabe, N., \& Smith, G.~P.\ 2010, MNRAS, 405, 2215 
\bibitem[Oguri et al.(2012)]{gravlens2} Oguri, M., Bayliss, M.~B., Dahle, H., et al.\ 2012, MNRAS, 420, 3213
\bibitem[Pasquali et al.(2009)]{pasquali} Pasquali, A., van den Bosch, F.~C., Mo, H.~J., Yang, X., \& Somerville, R.\ 2009, MNRAS, 394, 38 
\bibitem[Planck Collaboration~\etal(2011)]{sz3} Planck Collaboration, Ade, P.~A.~R., Aghanim, N., et al.\ 2011, A\&A, 536, A8 
\bibitem[Planck Collaboration et al.(2014)]{sz6} Planck Collaboration, Ade, P.~A.~R., Aghanim, N., et al.\ 2014, A\&A, 571, AA29
\bibitem[Piffaretti~\etal(2011)]{mcxc} Piffaretti, R., Arnaud, M., Pratt, G.~W., Pointecouteau, E., \& Melin, J.-B. \ 2011, A\&A, 534, A109
\bibitem[Popesso et al.(2004)]{popesso} Popesso, P., B{\"o}hringer, H., Brinkmann, J., Voges, W., \& York, D.~G.\ 2004, A\&A, 423, 449 
\bibitem[Puccetti et al.(2011)]{puccetti} Puccetti, S., Capalbi, M., Giommi, P., et al.\ 2011, A\&A, 528, A122
\bibitem[Refregier (2003)]{gravlens4} Refregier, A.\ 2003, ARA\&A, 41, 645
\bibitem[Reichardt et al.(2013)]{sz7} Reichardt, C.~L., Stalder, B., Bleem, L.~E., et al.\ 2013, \apj, 763, 127  
\bibitem[Richard et al.(2010)]{gravlens5} Richard, J., Smith, G.~P., Kneib, J.-P., et al.\ 2010, MNRAS, 404, 325 
\bibitem[Shan et al.(2015)]{m200c200} Shan, H., Kneib, J.-P., Li, R., et al.\ 2015, arXiv:1502.00313 
\bibitem[Skibba et al.(2011)]{skibba} Skibba, R.~A., van den Bosch, F.~C., Yang, X., et al.\ 2011, MNRAS, 410, 417
\bibitem[Sunyaev \& Zeldovich(1972)]{sz2} Sunyaev, R.~A., \& Zeldovich, Y.~B.\ 1972, Comments on Astrophysics and Space Physics, 4, 173 
\bibitem[Tinker et al.(2008)]{Tinker} Tinker, J., Kravtsov, A.~V., Klypin, A., et al.\ 2008, \apj, 688, 709 
\bibitem[Tundo et al.(2012)]{tundo} Tundo, E., Moretti, A., Tozzi, P., et al.\ 2012, A\&A, 547, A57 
\bibitem[Umetsu et al.(2014)]{gravlens7} Umetsu, K., Medezinski, E., Nonino, M., et al.\ 2014, \apj, 795, 163 
\bibitem[Valentinuzzi et al.(2011)]{red2} Valentinuzzi, T., Poggianti, B.~M., Fasano, G., et al.\ 2011, A\&A, 536, A34
\bibitem[van den Bosch et al.(2008)]{bosch} van den Bosch, F.~C., Aquino, D., Yang, X., et al.\ 2008, MNRAS, 387, 79 
\bibitem[Vikhlinin et al.(2009)]{vikhlinin} Vikhlinin, A., Burenin, R.~A., Ebeling, H., et al.\ 2009, \apj, 692, 1033
\bibitem[Voit(2005)]{voit} Voit, G.~M.\ 2005, Reviews of Modern Physics, 77, 207 
\bibitem[von~der~Linden~\etal(2007)]{bcg1} von der Linden, A., Best, P., Kauffmann, G., \& White, S.~D.~M.\ 2007, MNRAS, 379, 867
\bibitem[Weinberg et al.(2013)]{gravlens8} Weinberg, D.~H., Mortonson, M.~J., Eisenstein, D.~J., et al.\ 2013, PhR, 530, 87 
\bibitem[Weinmann et al.(2006)]{weinmann} Weinmann, S.~M., van den Bosch, F.~C., Yang, X., \& Mo, H.~J.\ 2006, MNRAS, 366, 2 
\bibitem[Wen et al.(2012)]{wen} Wen, Z.~L., Han, J.~L., \& Liu, F.~S.\ 2012, ApJS, 199, 34 
\bibitem[Wittman et al.(2006)]{gravlens9} Wittman, D., Dell'Antonio, I.~P., Hughes, J.~P., et al.\ 2006, \apj, 643, 128 

\end{thebibliography}
\end{document}